\documentstyle[bibnorm]{lamuphys}
\begin{document}

\title{Collapse Models}
\author{Philip Pearle}

\institute{Physics Department\\Hamilton College\\Clinton, New York 13323
\\ \\To be published in "Open Systems and Measurement in Relativistic Quantum Theory," 
F. Petruccione and H.P. Breuer eds. (Springer Verlag, 1999).}
\maketitle

\begin{abstract}
	This is a review of formalisms and models (nonrelativistic and relativistic) which 
modify Schr\"odinger's equation so that it describes wavefunction 
collapse as a dynamical physical process.  
	 
\end{abstract}

\section{Representing The World}\label{Representing}

	One of the major goals of theoretical physics is to 
find a useful mathematical object to represent the physical state of our world. 
In classical physics the object is a point in phase space. So far it has 
been possible to represent 
the world at an ``instant," i.e., on a spacelike 
hypersurface\footnote{One can imagine other 
possibilities, e.g., a representation might only be 
possible on the spacetime volume between two hypersurfaces; there might 
be another time parameter whose specification is also required; etc.}. One thing ``useful" means  
is that a dynamical equation can be found for the object, enabling predictions 
about the state of the world on a future spacelike hypersurface.  
  
	Is there such an object in quantum physics? 
In pursuing the research discussed here I have made some bets as 
to the nature of an eventually satisfactory physical theory. 
One of them is that there is such an object, 
a statevector in a suitable Hilbert space plus something \emph{more}. 
I shall argue that \emph{more} must be added because  
standard quantum theory (SQT) is a theory of choices without a chooser: 
\emph{more} is a chooser. Along the way I shall make a few opinionated remarks about 
other points of view.  
However, mostly I shall 
try to give a coherent account of rationale (section \ref{Representing}), 
formalisms (section \ref{Describing}), 
nonrelativisitic (including gravitational, section \ref{Nonrelativistic}) 
and relativisitic (section \ref{Relativistic}) models 
for a particular way of 
altering Schr\"odinger's equation so the statevector 
behaves in a manner consistent with this bet. 
 
	There is a point of view that the statevector does not 
represent the world but is only a mathematical tool 
enabling people to calculate the statistical 
outcomes of experiments\cite{Peres}. There is a point of view 
that only an ensemble of experiments is represented by the 
statevector\cite{Einstein,Pearle67,
Ballentine}. I disagree with both because I 
find it hard to believe such a successful theory as quantum theory 
either has no relevance to the unobserved world or to the 
individual world.  There is a 
point of view that it is the density matrix which 
is the proper object to represent the world.  There is a point of view that 
only ensembles of worlds can be represented.   
I shall comment on these later on in this section.

  I have of necessity been loose about what I mean 
by the part of ``the world" whose state the 
statevector represents.  
It should contain at least what we observe exists and  
what we thereby infer exists.  I will take as a minimal characterization that   
this means all particles and structures whose existence 
physicists at present generally agree upon.  These do not 
include the ``more" mentioned above, which is posited to be 
an additional structure in the world. However, it may 
be possible to include the ``more" in the statevector as well (see section 2.3h).  

	Essential is the view that 
there is a specifiable ``real world out there"\cite{BellGhost} 
whose behavior is completely 
consistent with our observations. Implied is that we are fair observers (what we 
see is \emph{there})  
and sound reasoners about the world so our conclusions can be trusted. \emph{Not} 
implied is that we are in any way essential to the 
description of the world.  Rather it is 
the theory of the world that is essential to the description 
of us and our observations (what has been called 
empirical reality\cite{d'espagnat}). I bet against 
the ultimate value of any 
theory whose logical consistency requires human beings\cite{Bellagainst}.   
 
	All statevectors having the same ``direction" in Hilbert space 
characterize the same state of the world. 
This generalizes the ``ray" (overall phase) 
independence of the SQT description to norm independence.     
In SQT the norm (squared) must be 1 because it represents 
the sum of probabilities, but 
that is not required when the statevector represents actuality. 
If two statevectors differ in direction by the slightest bit they 
represent two different states of the world. However,  
any one of a ``small" ball of directions will suffice to describe 
the same state of empirical reality\cite{Giancarloabner,Philabner}, i.e., 
an object may be empirically \emph{here} when a small amount of 
its wavefunction is elsewhere.    

	It is very possible that the statevector should represent more 
particles and structures than physicists now agree exist. However, I don't 
consider that the statevector 
should represent more   
variant replicas of our world, as in the many-worlds interpretation 
of the statevector\cite{Everett,DeWitt} because it is 
likely that the foundations of a coherent many-worlds picture requires 
human beings\cite{Euan,Lev,David}.   
But, even with 
such an interpretation, one of the statevectors  
in the many-worlds superposition describes our world, and it is this 
statevector I want to talk about. 

	The assumption   
that the statevector gives a \emph{complete} specification of 
the familiar particles and structures in the world 
excludes from discussion the deBroglie-Bohm 
model\cite{deBroglie, Bohm, PDH, Shelly, Valentini}.  
This model is akin to the models I discuss in that 
both contain an extra structure, a chooser, but the deBroglie-Bohm 
chooser is the positions of particles.  
I am betting that the statevector alone satisfactorily describes the particles
in the world.  Both the many-worlds 
and deBroglie-Bohm approaches to the central problem addressed  
here are interesting and have 
successes and difficulties, but I won't discuss them.  

	This central problem is arguably the most important unsolved 
problem in the foundations of quantum mechanics. It is 
brought about by certain features of the world as evidenced by 
our observations and inferences.  These are that 
\emph{events occur} in the world, that 
in certain circumstances they occur in a \emph{fundamentally 
unpredictable} (random) way, and that 
macroscopic objects (choose your own definition) are almost always 
highly \emph{spatially localized}---even 
objects whose location is an unpredictable event.  

	This latter behavior means that certain statevectors  
cannot represent possible states of the world, namely statevectors 
whose evolution describes a 
macroscopic object continually in a superposition  of \emph{here} and \emph{there} 
with comparably sized coefficients.   
The problem arises because, using  
the otherwise estimable SQT, one may start with 
a statevector representing a state of the world, in a 
situation such as a measurement in which
uncontrollable events occur, and it evolves via Schr\"odinger's equation 
into a statevector which does not represent a state of the world, 
but rather the sum of such states. 

	The founding fathers of quantum theory had a cure for 
this.  It is to modify the evolution of the statevector so that it instantly 
``collapses" (or ``jumps" or ``reduces") in such circumstances to a viable 
statevector representing 
the state of the world. The problem with this 
modification is that it is terribly ill-defined.  
It is supposed to be invoked whenever a measurement has been completed, but no one has 
been able to define what a measurement is or when completion is. 
Ad hoc means ``for this case only," 
and the prescription given is very ad hoc, with every situation 
requiring its own assessment. Perhaps due to the persuasive powers of 
John Bell\cite{Bellagainst}, more physicists nowadays think this collapse postulate is 
unsatisfactory.  

	There have been two approaches which can be said to follow in the 
footsteps of the founding fathers.  
	
	One approach is to try to make the instantaneous collapse postulate 
well defined without the need for ad hoc information.  This is my view of 
the ``consistent histories" program\cite{Griffiths,GellMannHartle,Omnes}. 
It has been described as a promissory note\cite{Saunders}   
which has serious problems in being fulfilled\cite{DowkerKent}.  The problem is that 
Nature selects a unique set of consistent collapse possibilities but the 
theory does not. 

	There is a test I 
think should be appled to all theories with fundamental pretensions\cite{PearleIt}.  
If, confronted with an initial statevector of 
a seriously complicated and realistic part of the world (e.g., the local galactic group), 
are there well-defined procedures for constructing the mathematical quantities which 
correspond to the possible real events which take place in the world? The point of this 
test is to see if a theory can do more than just 
handle some simplified models into which ad hoc features creep.  
So far the consistent histories approach does not pass this test. 

	This is as good a place as any to mention that the somewhat related 
``environmental decoherence" scheme\cite{Zeh, Zurek} does not pass this test either\cite{Squires}: 
so far the choice of ``environment" is ad hoc.
This scheme is a prime example of trying to use the density matrix to represent the world, which   
has the the following problems. If the density matrix is, 
in the Victorian characterization of 
d'Espagnat\cite{d'Espagnat2} ``Pure and Proper", i.e., $=|\psi,t>< \psi,t|$, then it 
is equivalent to a statevector and must undergo collapse in order to represent the 
world.  If it is Impure (mixed) and Proper then 
it may either represent an ensemble of worlds or our ignorance 
about our world, but not the world itself. But if it is
e.g., a density matrix traced over the ``environment," it  
begins as a Pure and Improper representation of our world and ends up interpreted as an  
Impure and Improper representation of an ensemble of worlds or of our ignorance 
about the world. The interpretation of a 
consistent theory should not undergo dynamical evolution. This is 
why I do not believe a density matrix is 
the appropriate object to represent the world.

\section{Describing Collapse}\label{Describing}  
	
	A second approach following the founding fathers, and espoused here, 
is to agree with them that 
the evolution of the statevector should be modified but to do it so the collapse is 
not instantaneous but, rather, follows a well-defined 
dynamics of a modified Schr\"odinger equation. This is a bet that there is 
a real physical process which causes events to occur which is not yet in physics  
and it is worthwhile to try to make a phenomenological model of it. 
Such a model is very strongly 
constrained because it must 
agree with empirical reality, which includes \emph{all} of 
tested physics, the random choices made by nature and 
the localization of macroscopic objects.
	 
	In order to describe the random choices 
made by nature the Schr\"odinger equation must have a chooser in it.  
The first such model\cite{Bohmbub} used some hidden variables.  Subsequently,  
I tried to use the phases of the 
superposition\cite{Pearle1} as chooser (which is appealing because 
nothing ``more" must be added beyond what is in the statevector).   
Then, betting 
that the chooser is one of many things in nature which fluctuate randomly,  
I settled on modelling it by external random noise\cite{Pearle2}, a choice 
which has been adopted in 
subsequent work\cite{Gisin, Diosi, Percival, Hughston, Fivel}. In this section I shall 
discuss variations on the theme of collapse formalism.

\subsection{Gambler's Ruin Game}\label{Gambler's}
	    
	With external random noise as the chooser, it turns out 
that the mechanism for obtaining 
agreement with the predictions of 
quantum theory is very simple\cite{Pearle5} 
which suggests one is on the right track. 
	
	The mechanism is the gambler's ruin game\cite{Feller}. Suppose one gambler has 
\$36 and another has \$64, and they toss a coin to determine who 
gives a dollar to the other.  Their dollar 
amounts fluctuate. Eventually  
one gambler wins all the money.  It is readily shown that the probability 
of winning all the money is .36 for the first (.64 for the second). 

	Precisely analogous 
is the modelled evolution of the initial statevector
 		
	\begin{equation}
	\label{E1}|\psi ,0>=.6|a_{1}>+.8|a_{2}>\,.
	\end{equation} 
	 
\noindent Under the influence of the random noise, the 
amplitudes multiplying $|a_{1}>$ and $|a_{2}>$ fluctuate.  
Eventually the statevector ends up as $|a_{1}>$ with 
probability .36 (ends up as $|a_{2}>$ with probability .64). This final result 
is just what the SQT collapse postulate accomplishes. 

\subsection{Simple Model}\label{Simple}  

	Here is a simplest example I can give of such a collapse dynamics. 
	
	Two equations are required for 
a collapse model.  The first is the dynamical 
equation which replaces Schr\"odinger's equation. Remarkably, it is 
possible to write this as a linear equation 
for the statevector\cite{Pearle89,Belavkin}. 
This simplicity is very useful e.g., it makes possible 
the relativistic collapse models discussed in section \ref{Relativistic}.  
The dynamical equation 
depends upon a classical white noise function w(t).  However, 
for this simple example, 
the solution actually depends only upon $\int_{0}^{t}w(t')dt'\equiv B(t)$ 
which is a classical Brownian motion function.  Here is the solution $|\psi ,t>_{B}$ 
when the initial state is (\ref{E1}): 

\begin{eqnarray} 
|\psi ,t>_{B}&&=
e^{-\frac{1}{4\lambda t}[B(t)-2\lambda tA]^{2}}|\psi ,0>\label{E2}\\
&&=.6|a_{1}>e^{-\frac{1}{4\lambda t}[B(t)-2\lambda ta_{1}]^{2}}
+.8|a_{2}>e^{-\frac{1}{4\lambda t}[B(t)-2\lambda ta_{2}]^{2}}\,.\label{E3}
\end{eqnarray}

\noindent  As will soon be seen, the operator $A$ 
determines the choices while $B(t)$ is the chooser.  
$A$'s eigenstates, $|a_{1}>$, $|a_{2}>$ (eigenvalues  $a_{1}$, $a_{2}$) are 
the states to which collapse occurs. $A$'s eigenvalue differences, 
together with $\lambda$ (a parameter of the theory)   
determine the collapse rate.  

	The evolution equation (\ref{E2}) tells us
what the initial statevector evolves into under a particular $B(t)$.  
The evolution is not unitary, 
so statevectors evolving under different $B(t)$'s have different norms.  
This plays a role in the second required equation, the probability rule.    
It gives the probability density 
for $B(t)$ to be the actual noise that occurs in nature:

\begin{eqnarray}
	{\cal{P}}_{t}\{B\}&&\equiv_{B}<\psi ,t|\psi ,t>_{B}\label{E4}\\
	&&=.36e^{-\frac{1}{2\lambda t}[B(t)-2\lambda ta_{1}]^{2}}
+.64e^{-\frac{1}{2\lambda t}[B(t)-2\lambda ta_{2}]^{2}}\,.\label{E5}
\end{eqnarray}

\noindent Eq. (\ref{E4}) says that the statevectors with the largest norm are the most 
likely to occur.  The total probability, according to (\ref{E5}), is
	 
\[\frac{1}{\sqrt {2\pi\lambda t}}
\int_{-\infty}^{-\infty}dB{\cal{P}}_{t}\{B\}=1\,.\]   

	To see how these equations work, from Eq. (\ref{E5}) we note that 
the most probable $B(t)$'s occur if 
$B(t)=2\lambda ta_{1}$ or $B(t)=2\lambda ta_{2}$ plus or minus a few  
standard deviations $(\lambda t)^{\frac{1}{2}}$. 
For small $t$ these regions overlap, but for 
large $t$ they don't.  
For example, set $B(t)=B_{0}(t)+2\lambda ta_{1}$, where the 
range of $B_{0}(t)$ is a small integer $\times(\lambda t)^{\frac{1}{2}}$. Then, 
for large $t$, Eqs. (\ref{E3}), (\ref{E5}) become  

\begin{eqnarray*}
|\psi ,t>_{B}&&\approx.6|a_{1}>e^{-\frac{1}{4\lambda t}B_{0}(t)^{2}}
+.8|a_{2}>e^{-\lambda t[a_{1}-a_{2}]^{2}}\\
{\cal{P}}_{t}\{B\}&&\approx
	.36e^{-\frac{1}{2\lambda t}B_{0}(t)^{2}}
+.64e^{-2\lambda t[a_{1}-a_{2}]^{2}}\,.
\end{eqnarray*}

\noindent Thus the probability associated with $B_{0}(t)$ in this 
range approaches .36 and the statevector approaches $\sim|a_{1}>$.  Of 
course, a similar argument holds for $B(t)=B_{0}(t)+2\lambda ta_{2}$, 
resulting in collapse to $|a_{2}>$ with probability .64. For other ranges of $B(t)$ 
the associated probability aproaches 0 for large $t$. This, then, is the way $B(t)$  
acts as chooser: it chooses either $2\lambda ta_{1}$ or $2\lambda ta_{2}$ to 
fluctuate around, and this determines the collapse outcome.  
	
	Collapse models such as this provide an explanation for the 
random results which occur in nature and are unexplained by SQT: the 
result was \emph{this} rather than \emph{that} because the noise fluctuated 
\emph{this way} rather than \emph{that way}.  What is left unexplained is at the 
next level: \emph{why} did the noise fluctuate
\emph{this way} rather than \emph{that way}?  A future theory may address 
this question by identifying the noise with something 
physical (e.g., gravitational fluctutions: see section \ref{Gravity}) and having dynamics for it.  
  
	It is useful to discuss various features of collapse dynamics in 
the context of this simple model before considering more sophisticated models.

\subsubsection{2.2a Density Matrix} 

	The density matrix describes the ensemble of statevectors which arise from all 
possible noises.  For our simple example it is

\begin{eqnarray}\rho(t)&&=\frac{1}{\sqrt {2\pi\lambda t}}
\int_{-\infty}^{\infty}{\cal{P}}_{t}\{B\}dB\frac{|\psi ,t>_{B}\,_{B}\negthinspace<\psi ,t|}
	{\,_{B}\negthinspace<\psi ,t|\psi ,t>_{B}}\nonumber\\
	&&=\frac{1}{\sqrt {2\pi\lambda t}}
\int_{-\infty}^{\infty}dB
|\psi ,t>_{B}\,_{B}\negthinspace<\psi ,t|\label{E6}\\
&&=.36|a_{1}><a_{1}|+.64|a_{2}><a_{2}|\nonumber\\
&&\qquad+.48\left[ |a_{1}><a_{2}|+|a_{2}><a_{1}|\right] e^{-\frac{\lambda t}{2}
(a_{1}-a_{2})^{2}}\label{E7}
\end{eqnarray}

\noindent which follows from (\ref{E4}), (\ref{E3}).  One clearly sees in the 
exponential decay of the off-diagonal density matrix elements that 
the collapse rate increases with increasing difference of $A$'s eigenvalues. 

	It should be emphasized that, while collapse 
dynamics gives rise to such density matrix behavior, it does not 
follow that such density matrix behavior implies collapse dynamics\cite{PearleDrexel}.  
Thus, the unitarily evolving statevector 

 \begin{eqnarray*}|\psi',t>_{B_{0}}&&=e^{-i{B_{0}}(t)A}|\psi,0>\\
 &&=.6|a_{1}>e^{-i{B_{0}}(t)a_{1}}
+.8|a_{2}>e^{-i{B_{0}}(t)a_{2}}
 \end{eqnarray*}
 
 \noindent where ${B_{0}}(t)$ has probability density 
 
\begin{eqnarray*}
{\cal{P}}_{t}\{B_{0}\}= e^{-\frac{B_{0}^{2}(t)}{2\lambda t}}
 \end{eqnarray*}
 
 \noindent has \emph{no} collapse behavior, just increasingly random phases,  
 yet at every instant 
 its density matrix is equal to (\ref{E7}):
 
 \begin{eqnarray*}
\rho '(t)= \frac{1}{\sqrt {2\pi\lambda t}}
\int_{-\infty}^{\infty}dB_{0}e^{-\frac{B_{0}^{2}(t)}{2\lambda t}}
|\psi',t>\negthinspace_{B_{0}}\,_{B_{0}}\negthinspace<\psi',t|=\rho (t)\,.
 \end{eqnarray*} 
 
 	It is often said that if the same density matrix 
 arises from such different ensembles of statevectors as 
 
 \begin{eqnarray*}&&\mbox{Ensemble A:}\quad |a_{1}>\mbox{ (probability }.36)\qquad
 	\mbox{plus}\qquad |a_{2}>\mbox{ (probability }.64)\\
 	&&\mbox{Ensemble B:}\quad .6|a_{1}>e^{i\theta_{1}}+.8|a_{2}>e^{i\theta_{2}}
 	\mbox{ (random }\theta_{1},\theta_{2})
 	\end{eqnarray*}
 	
 \noindent then the ensembles describe the same physics because they predict 
 (statistically) identical experimental outcomes. 
 The crucial point is that, 
 for this argument to hold, it is necessary for the theory to say that  
 experiments \emph{have} outcomes. 
 
 	Suppose the state (\ref{E1}) 
 represents the initial result of a measurement interaction where $|a_{1}>$, 
 $|a_{2}>$ describe macroscopically different apparatus states, 
 and suppose A, B 
 represent possible final apparatus ensembles.  
 Each member of ensemble A represents a specific measurement outcome and each member of 
 ensemble B does not. This conclusion is the same if a second apparatus 
 is introduced to 
 measure the state of the first apparatus and the two evolutions 
 take place as did the first (collapse for A, noncollapse for B).  Thus one 
 cannot make the argument cited above that ensembles A and B describe  
 the same physics because the argument depends upon both 
 ensembles having a collapse evolution, which B does not have. 
 
 	In a recent stimulating series of papers, Mermin implies that 
 an important lesson of quantum theory is that there can be no 
 physical difference if there is no experimental 
 difference\cite{Mermin}. This is in keeping 
 with the lesson learned by the previous generation of physicists 
 from the Ether issue.  Well, each succeeding generation may well  
 unlearn the lesson of the previous generation, e.g., when it's worth going to war. 
 The position taken here is the direct opposite, that an 
 important lesson of quantum theory is that just because we cannot 
 measure a difference does not mean that there is no difference.  After all, 
 why should \emph{we} be able to measure all that the world contains?  
 Thus, consider the 
 ensembles A and B but now take \emph{both} ensembles as subject to the 
 collapse dynamics (\ref{E2}), (\ref{E4}). Although the density matrix for both 
 ensembles remains $.36|a_{1}><a_{1}|+.64|a_{2}><a_{2}|$ during the evolution, 
 the dynamics recognizes that the states that make up A and B are different 
 and it treats them differently, leaving the states in A alone and collapsing the states in B.   
 
\subsubsection{2.2b Fourier Form}
	It is useful and interesting to write the statevector evolution (\ref{E2}) 
as a Fourier Transform:

\begin{equation}\label{E8} |\psi,t>_{B}=\frac{1}{\sqrt{2\pi}}
\int_{-\infty}^{\infty}d\eta e^{-\frac{\eta^{2}}{2}}
e^{i\eta \frac{1}{\sqrt{2\lambda t}}[B(t)-2\lambda tA]}|\psi,0> \, .
\end{equation}

\noindent  This says that the collapse 
evolution can be viewed as a gaussian-weighted sum of unitary evolutions.  
This is useful because unitary evolutions are well understood--for example, 
one may use Feynman diagram techniques with them (see section 4.2a).  It also 
may be viewed more profoundly, as a possible natural generalization of the 
usual unitary evolution of SQT. In SQT the statevector not only 
represents the world but it carries the extra burden of representing a 
probability, so its norm must be one and its evolution is limited to being unitary.  
When the statevector represents reality and  
is freed from representing a probability it 
is also freed to have a more general evolution than unitary.

	The density matrix can likewise be written as 
a Fourier transform using (\ref{E6})  and (\ref{E8}):

\begin{equation}\label{E9}\rho(t)=\frac{1}{\sqrt{\pi}}
\int_{-\infty}^{\infty}d\eta e^{-\eta^{2}}
e^{-i\eta \sqrt{2\lambda t}A}\rho(0)e^{i\eta \sqrt{2\lambda t}A}\, . 
\end{equation}

\noindent Again, as discussed in the previous section, 
although this is a collapse evolution, the Fourier form of the density matrix displays it as 
a Gaussian-weighted sum of non-collapse (unitary) evolutions.  

\subsubsection{2.2c Norm} 

	The statevector evolving according to (2) gets diminished in norm (e.g., from 1 to 
$\approx .6$ or $\approx .8$ in the example).  Further 
similar evolutions decrease 
the norm further.  However, as we have mentioned, the norm has no importance in a theory 
with collapse (e.g., amplitude ratios, operator eigenstates are norm-independent). 
Because the evolution (\ref{E2}) is linear in 
$|\psi>$, the right side may be multiplied by a function $F[t, B(t)]$ and 
the probability density (\ref{E4}) scaled by $F^{-2}$ without physical effect.  
For example, if $F=\exp[B^{2}(t)/4\lambda t]$, the evolution equation becomes 

\begin{eqnarray*}
|\psi,t>_{B}'= e^{B(t)A-\lambda t A^{2}}|\psi,0>
\end{eqnarray*}

\noindent so the final statevectors grow in norm 
(e.g., $\sim\exp \lambda t a_{i}^{2}$).  Incidentally, this may be written as 

\begin{eqnarray*}|\psi,t>'_{B}= 
e^{A\int_{0}^{t}dt[\frac{dB(t)}{dt}-\lambda A]}|\psi,0>\end{eqnarray*}

\noindent showing that (\ref{E2}) has a time-translation invariant form.
  
	If, because of prior experience with SQT, one is 
more comfortable with normalized eigenstates, 
one may feel free to normalize the statevector at any time, or even 
set $F=_{B}<\psi,t|\psi,t>_{B}^{-\frac{1}{2}}$ so that the statevector is 
normalized at every instant of time.  However, then manifest scale 
invariance is lost and the evolution equation is nonlinear in $|\psi>$.

	For evaluating collapse dynamics with a 
computer\cite{PercivalGisin} it is useful to have 
the statevector always of norm 1 since computers don't handle well numbers 
which get too small or large.  Also computers most easily generate 
random noise functions which average around 0, and also most easily 
calculate dynamical evolution in small steps.  To do this, one may  
consider evolution between $t$ and $t+dt$ and write the noise $dB(t)$ in this 
time interval as 

\begin{eqnarray*}dB(t)&&=dB_{0}(t)+2\lambda dt\bar A(t)\\
\bar A(t)&&\equiv\frac{\thinspace _{B}
\negthinspace<\psi,t|A|\psi,t>_{B}}{_{B}\negthinspace<\psi,t|\psi,t>_{B}} 
\end{eqnarray*}

\noindent thereby causing the probability density to 
be $\exp-dB_{0}^{2}(t)/2\lambda dt$, belonging to an increment 
of Brownian motion.  The statevector may be multiplied 
by an $F$ chosen to normalize it.  The result\cite{GPR, Gisin} is the nonlinear equation 

\begin{equation} d|\phi ,t>_{B_{0}}=
\left\{ dB_{0}(A-\bar A)-\lambda(A-\bar A)^{2}dt\right\}|\phi ,t>_{B}\,.\label{E10} 
\end{equation}

	In a recent paper\cite{GDS} the authors wrote that a linear evolution 
such as (\ref{E2}) is ``merely a mathematical relation" and went 
on to say that ``to be truly useful"  states should be normalized. 
Computational usefulness should not be confused with physical importance.  
The evolution (\ref{E2}) with initial state (\ref{E1}) 
models a physical quantity $dB/dt$ whose random fluctuations (both positive and negative)   
about one of the eigenvalues of $A$ drive the 
statevector toward the associated eigenvector.  
In the evolution (\ref{E10}) with initial state (\ref{E1}), 
the noise fluctuates about zero with positive 
excursions enhancing the eigenstate 
with the larger eigenvalue and negative  
excursions enhancing the eigenstate 
with the smaller eigenvalue. Indeed, although (\ref{E10}) is 
computationally equivalent to (\ref{E2}) and (\ref{E4}), 
one might argue that (\ref{E10}) is 
``merely a mathematical relation" since   
a symmetrical effect of positive and negative 
fluctuations about a mean value 
(such as is embodied in (\ref{E2})) is the usual type of 
behavior for a random physical quantity.   
As another example of the 
questionable physical nature of such descriptions as (\ref{E10}), 
it has been shown\cite{GG}
that such an equation can 
never lead to a relativistic collapse dynamics, 
unlike the situation with the linear equation (section \ref{Relativistic}).

\subsection{Less Simple Models}\label{Less}

	The model upon which more sophisticated collapse dynamics 
is based has the statevector evolution 

\begin{equation}\label{E11} |\psi,t>_{w}={\cal T}e^{-\frac{1}{4\lambda}
\int_{0}^{t}dt[w(t)-2\lambda A(t)]^{2}}|\psi,0>
\end{equation}

\noindent (${\cal T}$ is the time-ordering operator). This evolution is in what may be called 
the ``collapse interaction picture," 
where $A(t)\equiv \exp iHtA\exp -iHt$ is a Heisenberg operator 
and the statevector only evolves because of the collapse dynamics.  
The probability density for $w$ is 

\begin{equation}\label{E12} 
	{\cal{P}}_{t}\{w\}\equiv_{w}<\psi ,t|\psi ,t>_{w}\,.
\end{equation}  
  
	This is more complicated than the previous model because 
the statevector (\ref{E2})  
is a \emph{function} of $B(t)$ but the statevector 
(\ref{E11}) is a \emph{functional} of $w(t)$. 
The probability that this noise lies between $w(t)$ and $w(t)+dw$ is
$\prod_{t'=0}^{t}(2\pi\lambda/dt)^{-\frac{1}{2}}dw(t'){\cal P}_{t}\{w\}$: 
one imagines the time 
interval $t$ divided into infinitesimal segments $dt$, with each $w(t')$ 
as an independent variable. 

	However, the previous model is a special case of this, 
when $H=0$ so $A$ is time independent.   Then,  
writing $B(t)=\int_{0}^{t}dtw(t)$, (\ref{E11}) becomes 

\begin{eqnarray*} 
|\psi ,t>_{w}=e^{-\frac{1}{4\lambda}
[\int_{0}^{t}dtw^{2}(t)-\frac{B^{2}(t)}{t}]}
e^{-\frac{1}{4\lambda t}[B(t)-2\lambda tA]^{2}}|\psi ,0>\,.
\end{eqnarray*}

\noindent This is just the evolution (\ref{E2}) multipled by  
a time-dependent norm factor. Apart from this factor, the evolution 
only depends upon $B(t)$ and not upon $B$ at earlier times so one 
may integrate the probability density over $B$ at earlier times (using  
$w(t')=B(t')-B(t'-dt)$), obtaining the probability (\ref{E4}) of $B(t)$ alone.  

	One may think of (\ref{E11}) as describing collapse like (\ref{E2}) 
over each brief interval $dt$.  The equation attempts 
collapse to $A(t)$'s instantaneous eigenstates, with a rate proportional 
to the squared difference of its eigenvalues.  Since $A(t)$ is changing with time,   
its eigenvalues provide a ``moving target" for $w(t)$ to fluctuate around. 
Collapse to one state occurs if the collapse rate 
characterized by $\lambda$ is faster than the transition rate between these 
states characterized by $H$. Most generally, 
there is a competition between collapse and Hamiltonian dynamics. But, often states 
do not appreciably change during collapse, so the behavior is well approximated 
by setting $H=0$.

\subsubsection{2.3a Density Matrix} The density matrix which 
follows from (\ref{E11}), (\ref{E12}) is 

\begin{eqnarray}
\rho (t)&&=\int_{-\infty}^{\infty}\prod_{0}^{t}\frac{dw(t)}{\sqrt{2\pi\lambda/ dt}}
|\psi,t>_{w}\,_{w}\negthinspace\negthinspace<\psi,t|\label{E13}\\
&&={\cal T}e^{-\frac{\lambda}{2}\int_{0}^{t}dt
[A(t)\otimes 1-1\otimes A(t)]^{2}}\rho(0)\label{E14}
\end{eqnarray}

\noindent ($A\otimes B\rho\equiv A\rho B$ and ${\cal T}$ 
implies time ordering for operators to the left of $\rho(0)$ and 
time-reversed ordering for operators to the right).   
	
\subsubsection{2.3b Fourier Form} The 
Fourier transform of (\ref{E11}) is found 
by regarding each $w(t)$ as an independent variable 
with conjugate variable $\eta (t)$, resulting in 

\begin{eqnarray}\label{E15}
 |\psi,t>_{w}=\int_{-\infty}^{\infty} D\eta 
 e^{-\lambda\int_{0}^{t}dt\eta^{2}(t)}
 {\cal T}e^{i
\int_{0}^{t}dt\eta (t)[w(t)-2\lambda A(t)]}|\psi,0>
\end{eqnarray}

\noindent expressed as a superposition of unitary transformations.  

	The density matrix is easily expressed in 
Fourier form using (\ref{E13}), (\ref{E15}):

\begin{eqnarray}
 \rho(t)&&=\int_{0}^{t} D\eta 
 e^{-2\lambda\int_{0}^{t}dt\eta^{2}(t)}
 {\cal T}e^{-i2\lambda\int_{0}^{t}dt\eta (t) A(t)}\rho(0)
 {\cal T}_{R}e^{i2\lambda\int_{0}^{t}dt\eta (t) A(t)}\nonumber\\
 &&=\int_{0}^{t} D\eta 
 e^{-2\lambda\int_{0}^{t}dt\eta^{2}(t)}
 {\cal T}e^{-i2\lambda\int_{0}^{t}dt\eta (t)[A(t)\otimes 1-1\otimes A(t)]}\rho(0)
 \label{E16} 
\end{eqnarray}

\noindent (${\cal T}_{R}$ is the time-reversed ordering operator).  
Of course, if the $\eta$'s are integrated over, the result is (\ref{E14}).  
But, the form (\ref{E16}) displays the density matrix 
as a superposition of noncollapse unitary evolutions, each 
describing the interaction of the operator $A(t)$ with
Gaussian-weighted noise $\eta (t)$.  

\subsubsection{2.3c Many A's} To model the collapse of a system where states
 differ in a number of collapse-relevant features, each feature characterized by an operator 
 $A_{n}$, the evolution (\ref{E11}) may be generalized to
 
 \begin{equation}\label{E17} |\psi,t>_{w}={\cal T}e^{-\frac{1}{4\lambda}
\sum_{n}\int_{0}^{t}dt[w_{n}(t)-2\lambda A_{n}(t)]^{2}}|\psi,0>\,.
\end{equation} 

\noindent The $A_{n}$'s are usually taken to commute at equal times.  It is 
to their joint eigenstates that collapse tends, with a rate 
proportional to the sum of the squares of their eigenvalue differences.  

	As an example of how this works, suppose the $A_{n}(t)$'s are time independent 
and the initial state is $|\psi,0>=\sum_{i}c_{i}|i>$ where 
$|i>=|a^{(1)}_{i}>|a^{(2)}_{i}>$... and $A_{n}|a^{(n)}_{i}>=a^{(n)}_{i}|a^{(n)}_{i}>$.  
Then the analog of (\ref{E3}) follows from (\ref{E17}):  

\[ |\psi,t>_{B}=\sum_{i}c_{i}|i>e^{-\frac{1}{4\lambda}
\sum_{n}[B_{n}(t)-2\lambda t a^{(n)}_{i}]^{2}}\,.
\]

\noindent Collapse takes place to $|i>$ with probability 
$|c_{i}|^{2}$ when $B_{n}(t)\approx 2\lambda t a^{(n)}_{i}$. Then the magnitude of the 
jth eigenbasis vector decreases 
$\sim \exp-\lambda t \sum_{n}[a^{(n)}_{i}-a^{(n)}_{j}]^2$, showing that 
each different feature in $|i>$, $|j>$ increases the collapse rate.  

	An equivalent form of (\ref{E17}) may be obtained in terms of new operators 
$A'_{n}(t)$ and noise functions $w'_{n}(t)$ related to the old ones by 
a nonsingular matrix transformation $A_{n}(t)=\sum_{j}K_{nj}(t)A'_{j}(t)$, 
$w_{n}(t)=\sum_{j}K_{nj}(t)w'_{j}(t)$: 

 \begin{equation}\label{E18} |\psi,t>_{w'}={\cal T}e^{-\frac{1}{4\lambda}
\sum_{n}\int_{0}^{t}dt[w'_{i}(t)-2\lambda A'_{i}(t)]^{2}
G_{ij}[w'_{j}(t)-2\lambda A'_{j}(t)]^{2}}|\psi,0>
\end{equation} 

\noindent where the matrix $G=K^{T}K$ is symmetric and positive. 

	It is possible that one may choose $K$ to be singular.  For example, if $A_{1}$'s eigenvalues 
are all equal then $A_{1}$ and $w_{1}$ play no role in collapse. Then one may choose 
$K_{1j}=0$, so $G_{1j}=0$ and $A_{1}$ and $w_{1}$ are excised from (\ref{E18}).  (Of course, 
there is one less $w'$ to integrate over.)

\subsubsection{2.3d NonMarkovian Evolution} Another 
generalization\cite{PearleNoise} of the evolution (\ref{E11}) is 

\begin{equation}\label{E19} |\psi,t>_{w}={\cal T}e^{-\frac{1}{4\lambda}
\int\int_{0}^{t}dt_{1}dt_{2}[w(t_{1})-2\lambda A(t_{1})]^{2}
G(t_{1}-t_{2})[w(t_{2})-2\lambda A(t_{2})]^{2}}|\psi,0>\,.
\end{equation} 

\noindent This is a nonMarkovian evolution because it does not  
satisfy the usual Markov property that two evolutions, one from $t_{0}$ to $t_{1}$ 
followed by one from $t_{1}$ to $t$, are equivalent to a single 
evolution from $t_{0}$ to $t$.  In Eq. (\ref{E19}), 

\begin{equation}\label{E20} 
G(t_{1}-t_{2})=\frac{1}{2\pi}\int_{-\infty}^{\infty}d\omega e^{i\omega (t_{1}-t_{2})}\tilde G(\omega) 
\end{equation} 

\noindent where $G(t-t')$ is real, symmetric under exchange of $t$ and $t'$ 
and positive, i.e.,  $\int\negthinspace\negthinspace
\int_{\infty}^{\infty}dtdt'f(t)G(t-t')f(t')
\sim|\tilde f(\omega)|^{2}\tilde G(\omega)>0$ for arbitrary f(t),  
so $\tilde G(\omega)$ 
must be real, symmetric and positive. 
When $\tilde G(\omega)=1$ so $G(t_{1}-t_{2})=\delta (t_{1}-t_{2})$, (\ref{E19}) 
reduces to the Markovian (\ref{E11}).  A  reason 
for using the more general form (\ref{E19}) is to adjust the spectrum of $w$, 
which may be used 
to lower the amount of energy a system gets from $w$ during collapse.  

	To see how (\ref{E19}) gives rise to collapse, consider again the initial statevector 
(\ref{E1}), let $A$ be independent of $t$ and, for definiteness, 
choose $G(t-t')=(\alpha/2)\exp-\alpha|t-t'|$ 
($\tilde G(\omega)=\alpha^{2}/(\omega^{2}+\alpha^{2})$). Then 
 (\ref{E19}) yields 

\begin{eqnarray*}
&&|\psi,t>_{w}\sim \sum_{n=1}^{2}c_{n}(0)|a_{n}>e^
{\displaystyle -\frac{\{ B'(t)-2\lambda a_{n}[t-\alpha^{-1}(1-e^{-\alpha t})]\}^{2}}
{4\lambda [t-\alpha^{-1}(1-e^{-\alpha t})]}}\\
&&\mbox{where } B'(t)\equiv 
\int_{0}^{t}dt_{1}w(t_{1})[1-\frac{1}{2}(e^{-\alpha t_{1}}+e^{-\alpha(t- t_{1})}]\,. 
\end{eqnarray*}

\noindent For $t>>\alpha^{-1}$ this becomes (\ref{E3}), whose description of 
statevector collapse was discussed in section \ref{Simple}.

	In the Markovian case, the probability rule (\ref{E12}) applied 
at any time $T>t$ gives the same probability 
for the noise $w(t)$.  In the nonMarkovian case this is not so.  
In the present example 

\begin{eqnarray*}
{\cal{P}}_{T}\{w\}=\sum_{n=1}^{2}|c_{n}(0)|^{2}e^
{-\frac{1}{2\lambda}\int\int_{0}^{T}dt_{1}dt_{2}
[w(t_{1})-2\lambda  a_{n}]\frac{\alpha}{2}e^{-\alpha |t_{1}-t_{2}|}[w(t_{2})-2\lambda  a_{n}]}\,. 
\end{eqnarray*} 

\noindent  For fixed $t$, as $T$ increases past $t$, the 
probability of $w(t)$ only ``settles down" (becomes essentially independent of $T$) 
when $T-t>>\alpha^{-1}$.  

	What does this mean?  We interpret the probability as a measure of 
rational belief based upon present information\cite{Cox, Baierlein}. The future 
is not known at the present: 
present probability is conditional upon present time. Although the 
most we can ever know about $\{w\}$ is given by ${\cal{P}}_{\infty}$, at time $t$ 
we use ${\cal{P}}_{t}$ given by (\ref{E12}) because it represents all that can be known at time $t$.   
In the above example, if $\alpha^{-1}$ is small 
compared to the collapse time, the difference is not significant.   

\subsubsection{2.3e Fourier Form of the NonMarkovian Evolution}
The Fourier transformation representations of 
the statevector (\ref{E19}) and its associated density matrix are 

\begin{eqnarray}
 |\psi,T>_{w}&&=\int_{-\infty}^{\infty} D\eta 
 e^{-\lambda\int\int_{0}^{T}dtdt'\eta(t)G_{0,T}^{-1}(t,t')\eta(t')}\nonumber\\
 &&\qquad\qquad\qquad\cdot{\cal T}e^{i\int_{0}^{T}dt\eta (t)[w(t)-2\lambda A(t)]}|\psi,0>\label{E21}\\
\rho(T)&&=\int_{-\infty}^{\infty} D\eta 
e^{-2\lambda\int\int_{0}^{T}dtdt'\eta(t)G_{0,T}^{-1}(t,t')\eta(t')}\nonumber\\
 &&\qquad\qquad\qquad\cdot{\cal T}e^{-i2\lambda\int_{0}^{T}dt\eta (t)[A(t)\otimes 1-1\otimes A(t)]}\rho(0)\,.
 \label{E22} 
\end{eqnarray}

\noindent $G_{0,T}^{-1}(t_{1},t_{2})$ is the inverse of $G$ over the interval $(0,T)$:

\begin{equation}
 \int_{0}^{T}dt_{1}G(t,t_{1})G_{0,T}^{-1}(t_{1},t')=\delta (t-t')
 \label{E23} 
\end{equation}

\noindent and depends upon the differences of $T$, 0, $t_{1}$, $t_{2}$. For our example,  

 \begin{eqnarray*}
 G_{0,T}^{-1}(t,t')=\left(1-\frac{1}{\alpha^{2}}
 \frac{\partial^{2}}{\partial t^{2}}\right)\delta (t-t')
 &&-\frac{1}{\alpha^{2}}\{\delta (T-t)\delta '(T-t')+\delta (t)\delta '(t')\}\\ 
 &&-\frac{1}{\alpha}\{\delta (T-t)\delta (T-t')+\delta (t)\delta (t')\}\,. 
 \end{eqnarray*}

	The inverse of $G$ over an infinite interval, $\equiv G^{-1}(t-t')$, is much easier to 
find than its inverse over a finite interval.  Indeed, from (\ref{E20}) and (\ref{E23}) 
with the integral limits $(-\infty, \infty)$, it follows that $G^{-1}(t-t')$ is the 
Fourier transform of $1/\tilde {G}(\omega)$.

	Because $G^{-1}$ is so simple, it is useful to note that Eqs. (\ref{E21}), (\ref{E22}) 
may be written with $G_{0,T}^{-1}$ replaced by $G^{-1}$ and the double integral limits 
replaced by $(-\infty, \infty)$.  This can be seen from the identity

\begin{eqnarray*}
&&e^{-\frac{1}{2}\int\int_{-\infty}^{\infty}dtdt'f(t)G(t-t')f(t')}=\\
&&\qquad\qquad\int D\eta e^{-\frac{1}{2}\int\int_{-\infty}^{\infty}dtdt'\eta (t)G^{-1}(t-t')\eta (t')}
e^{i\int_{-\infty}^{\infty}\eta (t) f(t)}\,.
\end{eqnarray*}
 
\noindent With  $f(t)$ 
nonvanishing  only in the interval $0\leq t\leq T$ this becomes 

 \begin{eqnarray*}
&&e^{-\frac{1}{2}\int\int_{0}^{T}dtdt'f(t)G(t-t')f(t')}=\\
&&\qquad\qquad\int D\eta e^{-\frac{1}{2}\int\int_{-\infty}^{\infty}dtdt'\eta (t)G^{-1}(t-t')\eta (t')}
e^{i\int_{0}^{T}\eta (t) f(t)}
\end{eqnarray*}

\noindent which must be equivalent (manifest when the $\eta$'s are integrated over for 
time values outside the interval $(0,T)$ ) to 

\begin{eqnarray*}
&&e^{-\frac{1}{2}\int\int_{0}^{T}dtdt'f(t)G(t-t')f(t')}=\\
&&\qquad\qquad\int D\eta e^{-\frac{1}{2}\int\int_{0}^{T}dtdt'\eta (t)G_{0,T}^{-1}(t,t')\eta (t')}
e^{i\int_{0}^{T}\eta (t) f(t)}\,.
\end{eqnarray*}

	A generalized form of (\ref{E21}), also a Gaussian-weighted superposition of 
unitary transformations, is\cite{DiosiPearle}

\begin{eqnarray*}|\psi,T>_{w}&&=\int D\eta 
 e^{-\lambda\int\int_{-\infty}^{\infty}dtdt'\eta (t)G^{-1}(t-t')\eta (t')}\\
&&\qquad\cdot{\cal T}e^{i\int_{0}^{T}dt\eta (t)[w(t)-2\lambda A(t)]
 -i\int_{0}^{T}dtA(t)\int_{0}^{t}dt'F(t,t')w(t')}|\psi,0>\,.
 \end{eqnarray*}

\noindent In the additional $\eta$-independent term, $A$ 
interacts with the noise $w$ through an arbitrary function $F$: 
its asymmetrical time integrals are required to 
make the trace of the associated density matrix $=1$.  
This density matrix is readily shown to be

\begin{eqnarray*}
\rho(T)&&={\cal T}e^{-\frac{\lambda}{2}\int_{0}^{T}dtdt'
[A(t)\otimes 1-1\otimes A(t)]G'(t,t')[A(t')\otimes 1-1\otimes A(t')]}\\
&&\qquad \cdot  e^{-i\lambda \int_{0}^{T}dt[A(t)\otimes 1-1\otimes A(t)]\int_{0}^{t}dt'F(t,t')
[A(t')\otimes 1+1\otimes A(t')]}\rho(0) 
\end{eqnarray*}

\noindent ($G'(t,t')\equiv G(t-t')+\int_{0}^{t}dt_{1}\int_{0}^{t'}dt_{2} 
F(t,t_{1})G^{-1}(t_{1}-t_{2})F(t',t_{2})$). This is 
a special case of an even more general nonMarkovian 
density matrix first discussed by Strunz\cite{Strunz}, whose 
statevector evolution is based upon complex noise and allows nonHermitian operators.

\subsubsection{2.3f All-Time Dependent Evolution}\label{All} A variant of the nonMarkovian evolution 
(\ref{E19}) makes the statevector evolving from time $0$ to time $t$ nonetheless 
dependent upon $w$ for \emph{all} time:

\begin{equation}\label{E24} |\psi,t>_{w}={\cal T}e^{-\frac{1}{4\lambda}
\int\int_{-\infty}^{\infty}dt_{1}dt_{2}[w(t_{1})-2\lambda A_{0,T}(t_{1})]
G(t_{1}-t_{2})[w(t_{2})-2\lambda A_{0,T}(t_{2})]}\rho(0)
\end{equation}

\noindent where $A_{0,T}(t_{1})\equiv A(t_{1})$ for $0\leq t_{1}\leq t$ and $A_{0,T}(t_{1})\equiv 0$ 
elsewhere.  

	How is this to be interpreted?  If the evolution is Markovian 
($G(t_{1}-t_{2})=\delta (t_{1}-t_{2})$) then the statevector (\ref{E24}) 
differs from the statevector (\ref{E11}) only by an inessential numerical factor 

\[ e^{-\frac{1}{4\lambda}
\left\{ \int_{-\infty}^{0}dt_{1}w^{2}(t_{1})+\int_{t}^{\infty}dt_{1}w^{2}(t_{1})\right\}}\,. \]

\noindent i.e., although the expression for the statevector depends upon 
past and future $w$'s, these have no effect.  The interpretive issue lies 
with the probability rule (\ref{E12}).  The probability for $w(t)$ given by 
${\cal {P}}_{T}(w)$ changes quite abruptly as $T$ increases from less than $t$ 
to greater than $t$.  However, the interpretation of 
${\cal {P}}_{T}(w)$ is no different than already stated for 
a nonMarkovian evolution, namely it is a conditional probability, conditioned upon present information.  
At unaccessed times ${\cal {P}}_{T}(w)$ just gives a ``neutral probability estimate" of $w(t)$, 
 i.e., that of white noise with zero mean. 

	If the evolution is nonMarkovian, there is a new wrinkle.  
The statevector (\ref{E24}) and its associated  probability at time $T$ depend upon $w(t)$ 
in the future of $T$ and the past of $0$ in a non-neutral way, 
at least for a brief interval ($\approx \alpha^{-1}$ 
in our example) . There's nothing contradictory about this, 
it's just the way things are according to the model: at the statevector level  
the present depends at least a little bit on the future.  

	However, the evolution (\ref{E24}) gives the 
density matrix 

\begin{eqnarray} 
\rho (T)&&=\int_{-\infty}^{\infty} D\eta 
e^{-2\lambda\int\int_{-\infty}^{\infty}dt_{1}dt_{2}
\eta (t_{1})G^{-1}(t_{1}-t_{2})\eta(t_{2})}\nonumber\\
 &&\qquad\qquad\qquad\cdot{\cal T}e^
 {-i2\lambda\int_{0}^{T}dt\eta (t)[A(t)\otimes 1-1\otimes A(t)]}\rho(0)
 \label{E25} 
\end{eqnarray}

\noindent which is equal to the density matrix (\ref{E22}) 
(see explanation in previous section) which results from the evolution (\ref{E19}).   
Therefore, at the ensemble level, there is no effect of the future upon the present.

\subsubsection{2.3g Time-Smeared Evolution}
  
	Another nonMarkovian evolution form equivalent to (\ref{E24}) is suggested by the    
observation that, since $G$ is positive, we may construct $G^{\frac{1}{2}}$:

\[ 
G^{\frac{1}{2}}(t-t')\equiv\frac{1}{2\pi}\int_{-\infty}^{\infty}
d\omega e^{i\omega (t-t')}[\tilde G(\omega)]^{\frac{1}{2}}\space  
\] 

\noindent so $\int_{-\infty}^{\infty}dt_{1} 
G^{\frac{1}{2}}(t-t_{1})G^{\frac{1}{2}}(t_{1}-t')=G(t-t')$.  
For our example, $[\tilde G(\omega)]^{\frac{1}{2}}
=\alpha/(\omega^{2}+\alpha^{2})^{\frac{1}{2}}$  and 
$G^{\frac{1}{2}}(t-t')=(\alpha/\pi)K_{0}[\alpha(t-t')]$. 

	If $w'(t)\equiv\int dt G^{\frac{1}{2}}(t-t_{1})w(t_{1})$, 
$A_{s}(t)\equiv\int dt G^{\frac{1}{2}}(t-t_{1})A_{0,T}(t_{1})$ which are  
time-smeared variables  are 
introduced into (\ref{E24}) we obtain

\begin{eqnarray}\label{E26} |\psi,t>_{w}={\cal T}e^{-\frac{1}{4\lambda}
\int_{0}^{t}dt[w(t)-2\lambda A_{s}(t)]^{2}}|\psi,0>\,.
\end{eqnarray}

\noindent Although this looks Markovian, of course it is not. 
Because ${\cal T}$ time-orders $A(t)$, not $A_{s}(t)$, 
two sequential evolutions are not equivalent to one evolution since, for $t\geq t'$, 
$[{\cal T}A_{s}(t)][{\cal T}A_{s}(t')]\not= [{\cal T}A_{s}(t)A_{s}(t')]$. 
We remark that (\ref{E26}) may be useful 
in a relativistic model where spacelike smearing of an operator should 
be accompanied by timelike smearing.

\subsubsection{2.3h Quantizing the Noise}  Although it may appear that the following is not a 
collapse model, I shall argue that it is.  

	Consider an operator $W(t)$ to represent quantized noise\cite{Belavkin, HudsonParsarathy,
PearleQuantize}:

\begin{eqnarray}\label{E27} W(t)\equiv\sqrt{\frac{\lambda}{2\pi}}\int_{-\infty}^{\infty}d\omega
\left[ e^{-i\omega t}a(\omega)+e^{i\omega t}a^{\dagger}(\omega)\right]
\end{eqnarray}

\noindent where $a(\omega)$, $a^{\dagger}(\omega)$ are annihilation and creation 
operators ($[a(\omega),a^{\dagger}(\omega ')]=\delta (\omega-\omega ')$, etc.) 
of a mode of frequency $\omega$ which can take on positive \emph{or} negative values.  
This makes $W(t)$ unusual for a quantum field, which usually only has positive 
frequencies.  As a result, $[W(t),W(t')]=0$ which also is unusual for a quantum field.  
But it is a nice property for a quantum field which has classical aspirations 
since a complete eigenbasis for $W(t)$ is labelled by eigenvalues at \emph{each} time:

\begin{eqnarray*} W(t)|w>=w(t)|w>, 
\qquad <w|w'>=\prod_{t=-\infty}^{\infty}\delta[w(t)-w'(t)]\,.
\end{eqnarray*}

\noindent If $|0>$ is the state with no noise mode excited ($a(\omega)|0>=0$), 
one readily finds 

	\begin{eqnarray}\label{E28}  <w|0>=e^{-\frac{1}{4\lambda}\int_{-\infty}^{\infty}
	dtw^{2}(t)}\,.
\end{eqnarray}

	$dW(t)/dt$ commutes with $W(t)$: one must look elsewhere for $W$'s conjugate operator.  It is 

 \begin{eqnarray}\label{E29} \Pi(t)\equiv{\frac{i}{8\pi \lambda}}\int_{-\infty}^{\infty}d\omega
\left[- e^{-i\omega t}a(\omega)+e^{i\omega t}a^{\dagger}(\omega)\right]
\end{eqnarray}

\noindent where $[\Pi(t),\Pi(t')]=0$ and $[W(t),\Pi(t')]=i\delta(t-t')$.  

	Now, consider the unitary statevector evolution
	
\begin{eqnarray} |\Psi,T>&&={\cal T}e^{-i2\lambda\int_{0}^{T}dtA(t)\Pi(t)}|0>|\psi,0>\nonumber\\
&&=\int Dw|w>{\cal T}e^{-i2\lambda\int_{0}^{T}dt
A(t)\frac{1}{i}\frac{\delta}{\delta w(t)}}<w|0>|\psi,0>\nonumber\\
&&=\int Dw|w>{\cal T}e^{-\frac{1}{4\lambda}\int_{-\infty}^{\infty}dt
[w(t)-2\lambda A_{0,T}(t)]^{2}}|\psi,0>
\label{E30}
\end{eqnarray}

\noindent where the last step follows from (\ref{E28}).  Eq. (\ref{E30}) 
says $|\Psi,T>=\int Dw|w>|\psi,T>_{w}$, where $|\psi,t>_{w}$ is identical to  
(\ref{E24}) (Markovian version). The standard probability 
measure of the state $|w>|\psi,T>_{w}$ arising from (\ref{E30}) is equivalent to the 
probability rule (\ref{E12}). 

	In section \ref{Representing} 
I emphasized that such a superposition as (\ref{E30}) 
cannot represent a possible state of the world.  Then,  what is the 
meaning of (\ref{E30})?  

	In his famous ``cat paradox" paper\cite{Schrodinger}, Schr\"odinger wrote ``For each 
measurement one is required to to ascribe to the $\psi$--function (= the prediction 
catalog) a characteristic, quite sudden change\dots The abrupt change by measurement\dots  
is the most interesting part of the entire theory." (\ref{E30}) does not 
represent a possible state of the world but, in Schrodinger's terminology, it 
is the ``prediction catalog" of such states.  Each $|\psi,t>_{w}$ in the 
superposition represents a possible state of the world: one of these is 
realized, the rest are not.    Because of the 
dynamics presented here, an unambiguous reality assignment to one of the terms in 
the superposition (\ref{E30}) is allowed which 
is not possible with SQT.  Instead of measurement-induced (whatever that means) abrupt 
collapses (occurring at ambiguous times), \emph{collapses occur every $dt$ sec.}
Each $|\psi,t>_{w}$ ``branches" into a family of possible $|\psi,t+dt>_{w}$'s with 
the ``label state" $|w>$ distinguishing among them.  The states $|\psi,t>_{w}$ 
in (\ref{E30}) describe the choices and the attached noise state $|w>$ 
is the chooser. One may think of an Omar Khayyamish 
``moving finger" going from $t$ to $t+dt$ and choosing the actual $|w>$'s new $w(t+dt)$
along the way.  It is this feature which is 
either lacking or which is present in poorly defined form in interpretations of 
SQT without collapse dynamics.   

	Eq. (\ref{E30}) has a number of features which facilitate this interpretation.
\begin{enumerate}  
\item For those enamored of environmental decoherence, the noise is effectively a universal 
environment (which does not have to be defined on an ad hoc basis for 
each physical situation).  This is so desirable it suggests looking for a universal 
environment as the physical source for the noise (gravity is an attractive 
possibility---see section 3.1d).  \item The 
states $|\psi,t>_{w}|w>$ do not interfere with one another. This allows the interpretation 
that one of them corresponds to reality and the rest to unrealized choices  
(one would think that an unrealized choice should not influence reality).  \item The  
attached state $|w>$ acts as a complete label of the past history of the state 
$|\psi,t>_{w}$, making identification of its evolution unambiguous.  
\item	Each state $|\psi,t>_{w}$ is physically sensible in that 
macroscopic objects are (almost always) localized (see section 3.1a).  
\item The states $|w>|\psi,t>_{w}$ are always orthogonal because the $|w>$'s are 
always orthogonal.  
The states $|\psi,t>_{w}$ representing reality are not always orthogonal, e.g., 
if the initial state $|\psi,0>=\alpha|a_{1}>+\beta|a_{2}>$ evolves  
continuously under different $w$'s to $|a_{1}>$ or $|a_{2 }>$, these states obviously cannot 
be orthogonal during the evolution. (Indeed, it is my opinion that 
what has made a 
sensible SQT interpretation impossible is that the states 
which represent reality are not always orthogonal.) 
\end{enumerate}

	It is possible, in a future more sophisticated theory, 
that some of these attractive features may be altered, with attendant complications.  But, it is 
interesting that a collapse model based upon classical noise and a completely 
quantum model can be essentially equivalent.  Although I shall not pursue 
this any further here, a similar completely quantum 
construction can be made for the nonrelativistic and relativistic 
collapse models in the next sections.

\section{Nonrelativistic Collapse Models}\label{Nonrelativistic}

	The first Galilean-invariant collapse model was constructed by 
Ghirardi, Rimini and Weber (GRW)\cite{GRW1, Bell3}.  In their 
``Spontaneous Localization" (SL) model they introduced two important
concepts which may be regarded as connected to the two parameters 
which characterize their model. 

	The first parameter, 
$\lambda^{-1}\approx10^{16}$ sec $\approx 3\times 10^{8}$ yrs, accompanies the 
concept that collapse is caused by a physical process which acts on 
\emph{all} particles.  $\lambda$ is the collapse rate for a single particle, 
chosen small enough so that an individual particle in a superposition is 
negligibly affected, but large enough so that there is a dramatic effect 
on a macroscopic collection of $n$ particles in a superposed state, namely  
collapse in $(\lambda n)^{-1}$ sec.  

	The second parameter, the mesoscopic 
distance $a\approx 10^{-5}$cm, accompanies the concept that 
the collapse process narrows wavefunctions so that 
particles are spatially localized, but not localized too much.  A particle's 
widely spaced wavefunction is narrowed to dimensions $\approx a$ by collapse.  Since 
a narrowed wavefunction has increased energy by the uncertainty 
principle, if $a$ were too small then particles would get too 
much energy from the process (see section 3.1c).  If  
$a$ were too large, then macroscopic objects would not be well 
localized by the process.  

	I shall not discuss the SL model any further as it is not in one of the 
forms considered here.  It has a problem (i.e., 
the process destroys the symmetry of the wavefunction 
under particle exchange)  
and has been superseded by the CSL model discussed in the next section. (See 
however references \cite{GR,GP} for reviews of this seminal model.)

\subsection{CSL}

	In order to construct a Galilean-invariant model with features of SL 
in the form of a type we are considering, 
take the evolution equation (\ref{E17}) with many $A_{n}$'s and choose $n$ 
to be a spatial index so that $A_{n}(t)\rightarrow A({\bf x},t)$:

\begin{eqnarray}\label{E31} A({\bf x},t)\equiv 
	\frac{1}{(\pi a^{2})^{\frac{3}{4}}}\int d{\bf z}N({\bf z},t) 
	e^{-\frac{1}{2a^{2}}({\bf x}-{\bf z})^{2}}\,.
\end{eqnarray}

\noindent Essentially, $A({\bf x},t)$ is proportional to  
the number of particles/vol in a spherical volume 
of radius $a$ centered upon ${\bf x}$---say, 
for now, that the particles are nucleons.   In (\ref{E31}),  
$N({\bf z})\equiv \xi^{\dagger}({\bf z})\xi({\bf z})$ is the nucleon
number density operator,   
($\xi({\bf z})$, $\xi^{\dagger}({\bf z})$  are the nucleon annihilation and 
creation operators at ${\bf z}$) and 
$N({\bf z},t)=\exp iHt N({\bf z})\exp -iHt$ is the associated Heisenberg operator. 

	The statevector evolution  
	
\begin{eqnarray}\label{E32}|\psi,T>_{w}={\cal T}e^{-\frac{1}{4\lambda}\int_{T_{0}}^{T}dtd{\bf x}
	[w({\bf x},t)-2\lambda A({\bf x},t)]^2}|\psi,T_{0}>
\end{eqnarray}

\noindent and the probability rule (\ref{E12}) constitute what I called the 
``Continuous Spontaneous Localization" (CSL) model\cite{P89,GPR} 
(in appreciation of GRW's SL model).  In 
CSL the classical field $w({\bf x},t)$ chooses to fluctuate around 
the nucleon number density ($\times$ a constant, smeared over a volume $\sim a^{3}$) of one of the states  
in a superposition, thereby causing collapse to that state.

\subsubsection{3.1a Collapse Rate in CSL}  It remains to explore consequences of the model.  
In doing so it is useful to work with the density matrix 
 
\begin{eqnarray}\label{E33}
\rho (T)={\cal T}e^{-\frac{\lambda}{2}\int_{0}^{T}dtd{\bf x}
[A({\bf x},t)\otimes 1-1\otimes A({\bf x},t)]^{2}}\rho(0)\,.  
\end{eqnarray}
 
	Suppose that $|\psi,0>=\sum_{k}c_{k}(0)|n_{k}>$ where the normalized 
basis vectors $|n_{k}>$ correspond to different nucleon number density  distributions: 
$N({\bf x})|n_{k}>=n_{k}({\bf x})|n_{k}>$.  Setting $H=0$,
the off-diagonal elements of (\ref{E33}) decay as 

 \begin{eqnarray}\label{E34}
<n_{j}|\rho (T)|n_{k}>=c_{j}(0)c_{k}^{*}(0) e^{-\frac{\lambda T}{2}\int d{\bf x}
[a_{j}({\bf x})-a_{k}({\bf x})]^{2}}
\end{eqnarray}

\noindent where $a_{k}({\bf x})$ is given by (\ref{E31}) with $N({\bf z},t)$ 
replaced by $n_{k}({\bf z})$. 

	Consider $n$ particles in a \emph{small clump} of size $<<a$.  
Suppose the initial state is a superposition of 
such clumps with widely separated centers ${\bf x}_{k}$, 
where $|{\bf x}_{j}-{\bf x}_{k}|>>a$.   
We may make the approximation $n_{k}({\bf z})\approx n\delta ({\bf z}-{\bf x}_{k})$.   
Then (\ref{E31}), (\ref{E34}) give  

\begin{eqnarray*}
<n_{j}|\rho (T)|n_{k}>\approx c_{j}(0)c_{k}(0)^{*} e^{-\lambda T n^{2}}
\end{eqnarray*}	
 
\noindent showing that the rate of collapse to one clump is 
proportional to $n^{2}$.

	Now consider an 
\emph{extended homogeneous object} of uniform number density $n({\bf z})=\rho$,  
where the states $|n_{j}>$, $|n_{k}>$ describe the object in two 
different places.  Let their overlapping volume be $V_{0}$ and their nonoverlapping volume be $V_{1}$.  Then, 
by (\ref{E31}), 
$a_{j}({\bf x})\approx (4\pi)^{\frac{3}{4}}a^{\frac{3}{2}}\rho$ inside the volume of each.  Outside 
the volume of each, 
$a_{j}({\bf x})\approx 0$. The integral over the overlapping volume $V_{0}$ gives no contribution 
in (\ref{E34}), and we obtain 

\begin{eqnarray*}
<n_{j}|\rho (T)|n_{k}>\approx c_{j}(0)c_{k}^{*}(0) 
e^{-(4\pi)^{\frac{3}{2}}\lambda T (a^{3}\rho)(V_{1}\rho)}\,.
\end{eqnarray*}	
 
 \noindent Thus the collapse rate for such an extended object is proportional to
 the number $V_{1}\rho$ of nonoverlapping particles (and to the number of particles in a volume $\sim a^{3}$).
 Even if the object is small, a cube with sides $10^{-4}$ cm in length, with modest nucleon density 
 $\rho=10^{25}$ nucleons/cc it has a collapse rate of $\approx 10^{-8}$ sec.  This 
 is the reason for the assertion that macroscopic 
 objects are localized except for very brief intervals.
 
 	But, what is meant by localized?  
 A collapsed statevector, corresponding to an object considered to be 
 localized \emph{here}, generally has a ``tail." This is a ``small" piece of the wavefunction which 
 is not \emph{here} and which is generally decreasing exponentially with time 
 at the collapse rate (as evidenced by the above off-diagonal density matrix behavior). 
 For a discussion of criteria for the ``smallness" of 
 the tails see \cite{Giancarloabner, Philabner}.  
 
 	In CSL the collapse rate increases as superposed states differ more by having more particles in 
 different places.  In this way the interaction of a system with its 
 environment may increase the collapse rate\cite{Grigolini,BJK}. However, the prime initiator of the collapse 
 is the system itself when it is in a superposition 
 of spatially different states.
 
 	Some examples of superpositions have been given where the authors have \emph{felt} that 
 the superposed states are macroscopic enough so that collapse \emph{ought} to occur (e.g., a superposition 
 \emph{here}+\emph{there} of many photons\cite{AlbertVaidman}; a complex molecule 
 in a superposition of excited plus unexcited states\cite{HomeChat}) and criticized CSL for not predicting 
 collapse in these cases.  But it has been argued that there is no conflict 
 with experience in these cases because collapse indeed occurs when the full experimental 
 situation (e.g., observation of the photons by a human detector\cite{Aicardi}; 
 detection of the state of the complex molecule\cite{PearleSquiresLast}) 
 are taken into account: with CSL, opinion can be replaced by calculation.

 \subsubsection{3.1b Nonlocality and Locality in CSL} Collapse of an individual statevector is 
 a nonlocal process: something you do \emph{here} affects something \emph{there}.  For example, if 
 a particle is in a superposition of wavepackets \emph{here} and \emph{there} and you turn on a 
 position measuring device located near \emph{here}, then the wavepacket \emph{there} is affected. 
 Either the particle is detected so the wavepacket \emph{there} disappears or the 
 particle is not detected and the wavepacket \emph{there} is all that survives.  
 In either case the particle was neither \emph{here} nor \emph{there} before the 
 measurement and is either \emph{here} or \emph{there} after the measurement, so the 
 reality status of the particle \emph{there} has been changed by an action \emph{here}.
   
 	The model clearly displays accord with Bell's theorem: to produce agreement with the 
 predictions of quantum theory there is nonlocal influence.  The nonlocality arises in 
 two ways.  One is from the direct product structure of states which requires  
 the apparatus \emph{here} to be multiplied by the particle state \emph{there} so that 
 collapse engendered by the former affects the latter.  The other way is through the 
 probability rule which requires high probability $w({\bf x},t)$'s everywhere.  Thus 
 a high probability $w({\bf x},t)$ at the site of the apparatus is (nonlocally) correlated with 
 a  high probability $w({\bf x},t)$ at the site of the particle.
 
 	However, this nonlocality on the statevector level cannot be used to send 
 signals from \emph{here} to \emph{there} because one has no control over the 
 field $w({\bf x},t)$.  This is evident by considering the density matrix in
 Fourier form: 
 
  \begin{eqnarray*}
\rho (T)=\int D\eta e^{- 2\lambda\int_{-\infty}^{ \infty}dtd{\bf x}\eta^{2}({\bf x},t)} 
U_{\eta}(T)\rho (0)U_{\eta}^{\dagger}(T)
\end{eqnarray*}

\noindent where $U_{\eta}(T)={\cal T}\exp-i2\lambda \int_{0}^{T}dtd{\bf x}\eta	({\bf x},t)A({\bf x},t)$.  
As is well known, such a unitary transformation does not allow nonlocal influences, nor 
does a superposition of such.

\subsubsection{3.1c Experimental Tests of CSL}\label{Experimental}  
   Since CSL has different dynamics than SQT it provides testably different 
predictions.  A straightforward test might involve interference of a 
sufficiently large object\cite{Pearleexperiment,Zeilinger,Clauser}. For example, if an n-particle 
bound state wavefunction passes through a two-slit screen, the two emerging 
packets will fluctuate in amplitude (play the 
gambler's ruin game) according to CSL, so they will not have 
the same amplitude at a distance from the screen.  
The resulting interference pattern will therefore be different (diminished in contrast) 
from that predicted by SQT where the packets have the same amplitude.  
Such an experiment is difficult to do. 

	A more readily performed experiment is to measure 
the increased energy of particles due to the collapse process\cite{GR, Squires2}.  
The expectation value of the energy, $\bar H(T)\equiv\mbox{Trace}H\rho(T)$, can be found using   
(\ref{E33}).  The potential energy part of $H$ commutes with $A({\bf x},t)$ 
and, when the exponential in (\ref{E33}) is expanded in a 
power series in $\lambda$ the kinetic energy part of $H$ gives 
$\int d{\bf x}[A({\bf x},t),[A({\bf x},t),H]]=\mbox{const}\times\int d{\bf x}N({\bf x})$ so 
all terms of higher order than the first vanish.  The result for $n$ 
particles of mass $m_{p}$ is  
 
\begin{eqnarray*}
\bar H(T)=\bar H(0)+\frac{3}{4}\lambda T n \frac{\hbar^{2}}{2m_{p}a^{2}}\,.
\end{eqnarray*}   
  
\noindent This says $10^{24}$ nucleons gain an average of $\approx$.3 eV/sec which 
corresponds to a temperature increase of $\approx$ .001$^{\circ}$K.  Although the average 
energy increase is quite small, infrequently a particle suddenly gains a large amount of 
energy and that can be looked for.  

	We now note that different particles may have different collapse rates.  
To accomodate this we replace the particle number operator $N({\bf z},t)$ in (\ref{E31}) with 
$\sum_{\alpha}g_{\alpha}N_{\alpha}({\bf z},t)$, where $N_{\alpha}$ is 
the particle number operator for particles of type $\alpha$: 
a particle's collapse rate is $\lambda g_{\alpha}^{2}$, with $g_{e}$ for 
electrons, $g_{p}$ for protons, $g_{n}$ for neutrons.  

	Using (\ref{E33}), the probability/sec, $\Gamma$,  
of excitation of a bound state $|\psi>$ to an excited state $|\phi>$ 
(irrespective of the center of mass behavior) is found to be\cite{PearleSquires}

\begin{eqnarray*}
\Gamma=\frac{\lambda}{2a^{2}}|<\phi|\sum_{j}g_{\alpha (j)}({\bf x}_{j}-{\bf Q})|\psi>|^{2}
\end{eqnarray*}

\noindent to lowest order in (size of bound state/$a$)$^{2}$.  
Here the sum is over all particles in the bound state, ${\bf x}_{j}$ is the position 
coordinate of the $j$th particle and ${\bf Q}\equiv\sum_{j}m_{j}{\bf x}_{j}/\sum_{j}m_{j}$ 
is the center of mass operator.  This rate vanishes identically if 
$g_{\alpha(j)}\sim m_{j}$.

	A number of experiments currently being done for other purposes 
(e.g., dark matter searches, neutrino detection, proton decay, etc.) look for 
the sudden appearance of energy in a volume of matter.  To date, for purposes of 
testing CSL, the most sensitive experiment 
has been one which looks for X-rays appearing in a slab of Germanium\cite{Miley}. 
If a 1s electron in a Ge atom is ionized by the collapse process, 
the result is an X-ray pulse of magnitude 11.1 keV 
(the 1s electron's ionization energy, emitted as 
photons from the Ge ion decaying to its ground state) plus the kinetic 
energy of the ejected electron (converted to photons by the electron's 
collision with Ge atoms).  The excitation rate of 
such an electron is found from the equation above to have its largest value  
$\approx 5000[(g_{e}/g_{p})-(m_{e}/m_{p})]^{2}$ counts/keV-kg-day at $\approx 11.1$ keV, 
assuming that $a$ has the GRW value and that $\lambda$ has the GRW value for nucleons 
(so $g_{p}=g_{n}=1$)\cite{CPAN}. 

	The best experimental limit\cite{PRCA} is that the rate at  $\approx 11.1$ keV  
is less than $\approx .2$ counts/keV-kg-day, resulting in  
$g_{e}/g_{p}\leq 13 m_{e}/m_{p}$.  Thus,  according to CSL and experiment, 
electrons collapse much less rapidly than nucleons.  This is 
suggestive (but only suggestive, given the assumptions about the  
parameter values)  
of mass-proportional coupling, $g_{\alpha}\sim m_{\alpha}/m_{p}$.  This means that 
$N({\bf z},t)$ in (\ref{E31}) is 
to be replaced by $M({\bf z},t)/m_{p}$, where $M$ is the mass 
density operator.  Mass-proportional coupling suggests that 
collapse is connected to gravity.

\subsubsection{3.1d Collapse and Gravity}\label{Gravity}
 	
 	It is hoped that collapse dynamics will some day be seen to arise in a natural 
 way from another area of physics. 
 It has frequently been suggested that this area is gravity\cite{Karolyhazy, Penrose, 
 Diosi1, GGR, Pearlerel, PearleSquires2, Anandan, Percival, Hughston, Fivel}. 
 Failing gravity at present to move 
 toward collapse, collapse may move toward gravity as follows.  
 
 	Let us make a change of variables of type illustrated in (\ref{E18}).   
 We change from $A({\bf x},t)$ and $w({\bf x},t)$ to $M({\bf x},t)$ and $w'({\bf x},t)$:\footnote
 {Unlike (\ref{E18}), this transformation is singular because $G$ has 
 some vanishing eigenvalues:  its eigenvalue spectrum is 
 $\exp-a^{2}{\bf k}^{2}/2$ which vanishes at $|{\bf k}|=\infty$.  This can 
 be handled technically in various ways and, as discussed following 
 (\ref{E18}), causes no essential difficulty.}  
 
 \begin{eqnarray*}
A({\bf x},t)&&\equiv 
	\frac{1}{m_{p}(\pi a^{2})^{\frac{3}{4}}}\int d{\bf z}M({\bf z},t) 
	e^{-\frac{1}{2a^{2}}({\bf x}-{\bf z})^{2}}\\
w({\bf x},t)&&\equiv 
	\frac{2\lambda}{m_{p}(\pi a^{2})^{\frac{3}{4}}}\int d{\bf z}w'({\bf z},t) 
	e^{-\frac{1}{2a^{2}}({\bf x}-{\bf z})^{2}}\,.	
\end{eqnarray*}

\noindent Now, $\lambda a/c\approx 10^{-32}$ 
is a dimensionless number not far from 
$Gm_{p}^{2}/\hbar c\approx 10^{-38}$, as was pointed out by Diosi\cite{Diosi1}. 
By substituting for the new variables in Eq. (\ref{E32}) and 
replacing $\lambda$ by $Gm_{p}^{2}/\hbar a$ we obtain as evolution equation 

\begin{eqnarray}|\psi,T>_{w'}={\cal T}e^{-\frac{1}{\hbar}
\int_{T_{0}}^{T}dtd{\bf z}d{\bf z'}
	[w'({\bf z},t)-M({\bf z},t)] \frac{G}{a}
	e^{-\frac{({\bf z}-{\bf z}')^{2}}{4a^{2}}}[w'({\bf z'},t)-M({\bf z'},t)]}
	|\psi,T_{0}>\nonumber\\ \label{E35}	
\end{eqnarray}

 	We may think of (\ref{E35}) as the nonrelativistic limit of 
a General Relativistic collapse formulation, where 
M is replaced by the trace of the energy-momentum-stress tensor ($\div c^{2}$),  
and $w'$ is regarded as the curvature scalar ($\times c^{2}/8\pi G$). 
That is, the collapse is due to 
fluctuations in the curvature scalar about its 
``classical value," an eigenvalue of the operator representing the traced stress tensor. 	

	Karolyhazy \cite{Karolyhazy} was the first to suggest that 
fluctuations in the metric tensor could be related to collapse. Squires and I\cite{PearleSquires}  
gave a crude model for such fluctuations. In it, Planck mass ($m_{pl}$)  
point particles randomly appear and disappear 
from the vacuum in the neighborhood of 
e.g., a proton, remaining on average for a Planck time with an average 
density of one ``Planckon"  per 
proton Compton volume $(\hbar/m_{p}c)^{3}$.  It was shown that two quantities, 
the spectrum of fluctuations of the nonrelativistic limit of 
the curvature scalar ($\times G^{-1}$), $w\equiv(4\pi G)^{-1}\nabla ^{2}\phi$,  
($\phi$ is the gravitational potential) and the energy increase of 
the proton (due to the random gravitational force $-m_{p}\nabla\phi$ exerted 
by the Planckons), match the comparable 
CSL expressions with certain numerical coefficients, so that two equations were obtained 
for the two CSL parameters:

 \begin{eqnarray*}
\lambda  =\frac{1}{2(3\pi)^{\frac{1}{2}}}\frac{Gm_{p}^{2}}{\hbar a}
\mbox{,\qquad}a=\left(\frac{3}{\pi^{2}}\right)^{\frac{1}{4}}\frac{\hbar}{4m_{p} c}\sqrt{\frac{m_{pl}}{m_{p}}}
\approx 1.4\times 10^{-5}\mbox{cm}\,.
\end{eqnarray*}

\noindent Although this model should not be taken very seriously, it does suggest that features of 
CSL might arise from an understanding of curvature fluctuations.  

	Gravitional quantities other than curvature may be considered to 
give rise to collapse. Diosi\cite{Diosi1} was the first to propose a 
gravitational collapse model of the form discussed here, with fluctuations in the 
scalar potential about its ``classical value" causing collapse.    
His evolution equation  may be written as 
   
\begin{eqnarray}&&|\psi,T>_{w}={\cal T}e^{-\frac{1}{\hbar}
\int_{T_{0}}^{T}dtd{\bf x}d{\bf x'}
	[w({\bf x},t)-\phi({\bf x},t)]\frac{\nabla_{x}\cdot\nabla_{x'}\delta({\bf x}-{\bf x}')}{G}
	[w({\bf x}',t)-\phi({\bf x}',t)]}
	|\psi,T_{0}>\nonumber\\ \label{E36}\\
	&&\phi({\bf x},t)\equiv G\int\frac{d{\bf z}}{|{\bf x}-{\bf z}|}M_{a}({\bf z},t),\mbox{\space} 
	M_{a}({\bf z},t)\equiv \int d{\bf z}'\frac{1}{(2\pi a^{2})^{\frac{3}{2}}}
	e^{-\frac{({\bf z}'-{\bf z})^{2}}{2a^{2}}}M({\bf z}',t)\nonumber
\end{eqnarray}

\noindent It was pointed out by Ghirardi, Grassi and Rimini\cite{GGR} that $M_{a}({\bf z},t)$, \emph{not}  
$M({\bf z},t)$, must be used in the expression for $\phi({\bf x},t)$.  If $M({\bf z},t)$ is used, 
$a$ is effectively  equal to the particle's Compton wavelength 
which means that particles get too much energy due to collapse. 

	 With a suitable change of 
variable, $w({\bf x},t)=G\int d{\bf z}w'({\bf z},t)/|{\bf x}-{\bf z}|$, 
(\ref{E36}) can look much like (\ref{E35}).  
Using $\nabla_{x}\cdot\nabla_{x'}=-\nabla_{x}^{2}$ in the integrand of (36) and integration  
by parts one obtains 

\begin{eqnarray*}|\psi,T>_{w}={\cal T}e^{-\frac{4\pi}{\hbar}
\int_{T_{0}}^{T}dtd{\bf x}d{\bf x'}
	[w'({\bf x},t)-M_{a}({\bf x},t)]\frac{G}{|{\bf x}-{\bf x}'|}
	[w'({\bf x}',t)-M_{a}({\bf x}',t)]}
	|\psi,T_{0}>	
\end{eqnarray*}

\noindent which has a similar interpretation to (\ref{E35}). 

	Penrose's suggestion\cite{Penrose} that collapse arises from the gravitational energy 
associated with the difference of gravitational fields of superposed states may be implemented by  

\begin{eqnarray}\label{E37}|\psi,T>_{w}={\cal T}e^{-\frac{1}{\hbar G}
\int_{T_{0}}^{T}dtd{\bf x}d{\bf x'}
	[{\bf w}({\bf x},t)-\nabla \phi({\bf x},t)]^{2}}|\psi,T_{0}>	
\end{eqnarray} 

\noindent where fluctuations in the gravitational field ${\bf w}\equiv{\bf g}$ 
about its ``classical value" causes collapse. 

	The density matrices associated with the evolutions (\ref{E35}), (\ref{E36}) and (\ref{E37}) 
are respectively 
  
\begin{eqnarray}
\rho (T)={\cal T}e^{-\frac{1}{2\hbar}\int_{0}^{T}dtd{\bf z}d{\bf z}'\Delta M({\bf z},t)
\frac{G}{a}
e^{-\frac{({\bf z}-{\bf z}')^{2}}{4a^{2}}}
\Delta M({\bf z}',t)}\rho(0)\label{E38}	
\end{eqnarray}

\noindent where $\Delta M({\bf z},t)\equiv[M({\bf z},t)\otimes 1-1\otimes M({\bf z},t)]$, and 

\begin{eqnarray}
\rho (T)={\cal T}e^{-\frac{1}{2\hbar}\int_{0}^{T}dtd{\bf z}d{\bf z}'\Delta \phi({\bf z},t)
\frac{\nabla_{z}\cdot\nabla_{z'}\delta({\bf z}-{\bf z}')}{G}
\Delta \phi({\bf z}',t)}\rho(0)\label{E39}	
\end{eqnarray}

\noindent where $\Delta \phi({\bf z},t)\equiv[\phi({\bf z},t)\otimes 1-1\otimes \phi({\bf z},t)]$, 
and 

\begin{eqnarray}
\rho (T)={\cal T}e^{-\frac{1}{2\hbar G}\int_{0}^{T}dtd{\bf z}
[\Delta \nabla\phi({\bf z},t)]^{2}}\rho(0)\label{E40}	
\end{eqnarray}

\noindent where 
$\Delta \nabla\phi({\bf z},t)\equiv[\nabla \phi({\bf z},t)\otimes 1-1\otimes\nabla \phi({\bf z},t)]$.
Eqs. (\ref{E39}) and (\ref{E40}) are the same, as may be seen by 
integrating by parts in (\ref{E39}).  For purposes of comparison with (\ref{E38}) 
they also may be written as 

\begin{eqnarray}
\rho (T)={\cal T}e^{-\frac{1}{2\hbar}\int_{0}^{T}dtd{\bf z}d{\bf z}'\Delta M({\bf z},t)
\frac{G}{|{\bf z}-{\bf z}'|}
[\int_{0}^{|{\bf z}-{\bf z}'|}\frac{ds}{\sqrt{\pi a^{2}}}e^{-\frac{s^{2}}{4a^{2}}}]
\Delta M({\bf z}',t)}\rho(0)\,.\label{E41}	
\end{eqnarray}

	Consider the collapse rate ($\times\hbar$) for a superposition of 
two states of different spread out $>>a$ mass distributions.  
According to (\ref{E38}), this is essentially $\sum G(\Delta M)^{2}/a$, 
where the sum is over 
cells of volume $a^{3}$	into which space is divided, and $\Delta M$ is the difference of 
the state's masses in each cell, i.e.,  the sum of the 
gravitational mass-difference self-energy of each cell. According to (\ref{E41}), 
the collapse rate ($\times\hbar$) is essentially the 
gravitational energy associated with the mass-difference (since the integral over $s\approx 1$). 
The collapse rate for (\ref{E38}) is, so to speak, 
a local gravitational energy and the collapse rate for (\ref{E41}) is a global gravitational energy.  

	Anandan\cite{Anandan} has suggested that a natural generalization of Penrose's 
suggestion to General Relativity is that collapse arises because of differences in 
the connection $\Gamma_{\mu\nu}^{\rho}$.  Unfortunately, $[\Delta \Gamma]^{2}$ is 
not positive definite so this suggestion cannot be implemented with the formalism 
presented here. 

	There have been other proposals regarding gravitational collapse. Collapse 
based upon energy differences (\emph{not} energy density differences) was first 
proposed by Bedford and Wang\cite{Bedford and Wang}. It has received an elegant formulation 
by Hughston\cite{Hughston} which, however, is equivalent to (\ref{E11}) with $A(t)=H$ and 
$\lambda^{-1}=\hbar \mu_{pl}c^{2}$. The problem with such a proposal is 
that, in collapse as in real estate, location is everything and 
energy localization does not lead to spatial localization. The proposal of Percival\cite{Percival} 
is based upon energy localization for small distances: it's extension to large distances has 
not been completed.  The proposal of Fivel\cite{Fivel} 
is based upon collapse acting to decrease a measure of entanglement and has also not been 
completed. One cannot say at present whether these two models  
localize satisfactorily.

\section{Relativistic Collapse Models}\label{Relativistic}

	One reason for trying to construct a special relativistically invariant collapse 
model is that, like Mount Everest, the symmetry is \emph{there}.  But there are other reasons as well. 
It seems that collapse and relativity are intimately related.  In  
nonrelativistic quantum theory, one cannot communicate to \emph{there} by initiating 
collapse \emph{here}, for any experimental setup whatsoever.  This is remarkable:  
why should a nonrelativistic theory be prevented from long-distance communication in this way when 
one can communicate long-distance another way, by just sending a particle from \emph{here} 
to \emph{there} with sufficient velocity?  It looks like this noncommunicability is 
a holdover from a relativistic theory where long distance (superluminal) communication 
could make it difficult to have a consistent theory.

	It might be possible to have collapse in a preferred reference frame, 
i.e., the comoving frame of the universe, and yet have the experimental results (but not the theoretical structure) 
agree with relativity.  Aharonov and Albert\cite{AA} have shown that an instantaneous 
collapse postulate in a preferred frame will not predict 
experimental results which disagree with special relativity. This assumes that collapse produces 
well defined and sensible states, 
which so far has not been achieved by any instantaneous collapse postulate but has been achieved 
by collapse models.  However, collapse models also bring along with them other effects, and these may 
not be relativistically invariant in a preferred-frame collapse model.  For example, the spontaneous 
excitation rate of atoms in CSL 
can be regarded as a clock whose ticking is a measure of the collapse rate.\footnote{This was 
mentioned to me some time ago by Renata Grassi. It 
suggests that the collapse rate of any moving object is time dilated---see section 4.1d}  
Then a moving atom's rate will be time-dilated and one may discern the special frame as the 
one in which the rate is fastest. 

	Of course there are many phenomena which are preferred frame dependent, e.g., the 2.7$^{\circ} K$ 
blackbody radiation.  There is nothing intrinsically wrong with having collapse produce 
preferred frame phenomena.  Still, it would be rather peculiar (although the universe \emph{is} peculiar) 
for collapse, which allows quantum theory to be compatible with 
relativity, to carry along with it phenomena which 
are not relativistically invariant. So, it is not unreasonable to 
guess that collapse may be completely relativistically 
invariant and to construct relativistic collapse models.

\subsection{RCSL1}\label{RCSL1: First Relativistic Collapse Model}

	The first relativistic collapse model\cite{Pearlerel, GGP} has a Markovian evolution: 
	
\begin{eqnarray}\label{E42}|\psi,\sigma>_{w}={\cal T}e^{-\frac{1}{4\gamma}\int_{\sigma_{0}}^{\sigma}dx
	[w(x)-2\gamma \phi(x)]^2}|\psi,\sigma_{0}>\,.
\end{eqnarray}	

\noindent ($\sigma_{0}$, $\sigma$, represent spacelike hypersurfaces, 
$x\equiv ({\bf x},t)$, $dx\equiv d{\bf x}dt$ and $\phi$ is a Heisenberg scalar relativistic quantum field). 
Here collapse occurs because of differences of scalar field amplitude (not because of 
differences of particle number density as in (\ref{E32})).  
The reason for 
the scalar field is to obtain the parameter $a$ in a natural relativistic way. $a$ 
is the Compton wavelength corresponding to $\phi$'s	mass $\mu=\hbar/ac\approx1$eV.  
The Hamiltonian under which $\phi$ evolves includes the interaction 
$g\phi :\negthinspace\negthinspace\bar{\psi}\psi\negthinspace\negthinspace :$ of a fermion field with $\phi$ 
so that $\phi$ ``dresses" a fermion, i.e., surrounds it with an 
average Yukawa field of extension $a$.  

	Collapse described by Eq. (\ref{E42}) works as follows.   
A Fermion in a superposition of \emph{here} plus \emph{there} is entangled with its scalar field: 
$|\psi,\sigma_{0}>=|$Fermion \emph{here}$>$$|$scalar \emph{here}$>$ + $|$Fermion \emph{there}$>$$|$scalar \emph{there}$>$. 
$w$ ``chooses" to fluctuate around 
one of these scalar fields causing collapse to its state and, in so doing, it collapses the attached Fermion 
state as well. The collapse rate of a massive Fermion in a superposition of 
locations ${\bf x}_{1}$, ${\bf x}_{2}$ is\cite{Pearlerel}:

\begin{eqnarray*}\frac{\gamma g^{2}}{2}\int d{\bf x}_{1}
\left[ \frac{e^{-\frac{|{\bf x}-{\bf x}_{1}|}{a}}}{4\pi |{\bf x}-{\bf x}_{1}|}-
\frac{e^{-\frac{|{\bf x}'-{\bf x}_{1}|}{a}}}{4\pi |{\bf x}'-{\bf x}_{1}|}\right]^{2}=
\frac{\gamma g^{2}a}{16\pi}\left[1- e^{-\frac{|{\bf x}-{\bf x}'|}{a}}\right]\,.
\end{eqnarray*}	 

\noindent In the comparable CSL expression, $\gamma g^{2}a/16\pi\rightarrow \lambda$ and 
$\exp-|{\bf x}-{\bf x}'|/a\rightarrow \exp-({\bf x}-{\bf x}')^{2}/2a$.  

	This is the good news: collapse works well.

\subsubsection{4.1a Vacuum Excitation} The bad news is 
that collapse works \emph{too} well. For simplicity, remove the Fermions from the model. 
The $\phi$-vacuum state may be written as 
a superposition of eigenstates of $\phi(x)$.  
The evolution (\ref{E42}) acting on the vacuum proceeds to 
collapse it toward one of these eigenstates.  
This amounts to excitation of the vacuum: $\phi$-particles are created.  
Each momentum ${\bf k}$ mode of the vacuum may be regarded 
as a harmonic oscillator in its 
ground state. It gets excited so that its average energy 
increases as $\bar{H}=\frac{1}{2}\mu\lambda t$, 
a modest rate of $\approx$1 eV/300 million years. 
 However there are an infinite number of modes so the total average energy 
increases as 

\begin{eqnarray*}\bar{H}=\frac{\lambda\mu t}{2}\sum_{\mbox{modes}}1=\frac{\lambda\mu t}{2}\frac{V}{(2\pi)^{3}}\int d{\bf k}
\end{eqnarray*}

\noindent i.e., at an infinite energy/sec-vol rate.  When Fermions are restored to the model they also are produced 
at an infinite energy/sec-vol rate.

	In a relativistic theory, when \emph{any} vacuum excitation occurs it must be infinite 
excitation. (A particle of a particular four-momentum produced from the vacuum in one frame 
has a different momentum in another frame so, since all frames are equivalent, all 
frames must have particles of all momenta come out of the vacuum.) Therefore, to make a sensible relativistic theory  
\emph{all} vacuum excitation must be eliminated.

 \subsubsection{4.1b Pictures of Relativistic Collapse}
 
 	Before looking at the cause of vacuum excitation and exploring how to eliminate it, I want to 
briefly mention a few features of the RCSL1 collapse process.   
The elucidation of these and other features is mainly due 
to Ghirardi and coworkers\cite{GGP, GG, G2,Gvol} (see
his contribution in this volume). The features mentioned here turn out to 
embody a proposal of Aharonov and Albert\cite{Aharonov and Albert1, BreuerPetruccione} 
regarding instantaneous collapse in a (noninstantaneous) collapse model.   

	Essentially, in each reference frame the collapse process works as 
in nonrelativistic CSL (except that the collapse time 
for a moving object is time-dilated---see section 4.2d).  
This means that the picture of collapse dynamics is frame dependent.  
Suppose a particle is in a superposition 
of wavepackets \emph{here} and \emph{there} and that apparatuses 
\emph{here} and \emph{there} are set up 
to detect the particle at the same time in a particular reference frame.  In this frame 
at that time both apparatuses simultaneously 
trigger collapse: say, the packet \emph{here} grows inside the apparatus \emph{here} while the 
packet \emph{there} fades away inside the apparatus \emph{there}.  
However, in any other reference frame,   
only one of the apparatuses 
triggers the collapse and the packet not in it grows or fades before it 
reaches the other apparatus.  Thus, unlike CSL, in RCSL 
\emph{what} triggers collapse and \emph{when} or \emph{where} 
the collapse takes place are frame-dependent.  
The model does not give objective (observer-independent) answers to \emph{what}, \emph{when} and 
\emph{where} but, of course, these are not measureable properties. 

\subsection{Removing Vacuum Excitation in Lowest Order: the Tachyon}\label{Removing}
	
	To see the reason for the infinite vacuum excitation, and how to cure it, 
we begin by generalizing the evolution (\ref{E42}), using a combination of 
the generalizations (\ref{E18}) and (\ref{E19}):\footnote{
For a more detailed presentation of the material in this and the remaining sections, see  
reference \cite{PearleRCOL}.  In what follows we set $\hbar=c=1$ unless mentioned otherwise.} 

\begin{equation}\label{E43}|\psi ,\sigma>={\cal T} e^{-{1\over 4\gamma}\int_{\sigma_{0}}^{\sigma}dxdx'
   			[w(x)-2\gamma \phi(x)]G(x-x')[w(x')-2\gamma \phi(x')]}|\psi ,\sigma_{0}>
\end{equation}

\begin{equation}
\mbox{where }G(x-x')={1\over (2\pi)^{4}}\int dke^{ik\cdot(x-x')}\tilde G(k^{2})\,.\label{E44}\end{equation}
	
\noindent The nonMarkovian evolution (\ref{E43}) reduces to the Markovian evolution 
(\ref{E42}) when $\tilde G(k^{2})=1$ ($G(x-x')=\delta(x-x')$). 

	The density matrix associated with (\ref{E43}), in Fourier form, is 
	
	\begin{eqnarray}\rho (\sigma)&&={\cal T}\int D\eta e^
{-2\gamma\int_{-\infty}^{\infty} dxdx'\eta(x)G^{-1}(x-x')\eta(x')}\nonumber\\
&&\qquad\cdot e^{-i2\gamma\int_{\sigma_{0}}^{\sigma}dx\eta(x)\phi(x)}\rho (\sigma_{0})
e^{i2\gamma\int_{\sigma_{0}}^{\sigma}dx\eta(x)\phi(x)}\,.
\label{E45}
\end{eqnarray}

\noindent We see from (\ref{E45}) that the density matrix is the same as for a non-collapse 
unitary evolution where the scalar field $\phi(x)$ interacts with a classical noise 
$\eta(x)$ with correlation function $\overline{\eta(x)\eta(x')}=(4\gamma)^{-1}G(x-x')$ and 
spectrum $(4\gamma)^{-1}\tilde G(k^{2})$.  

	As is well known, noise will cause a transition between two states if
it supplies a four-momentum equal to their four-momenta difference.  To create a particle
out of the vacuum, the noise has to supply the four-momentum 
of the particle. For the Markovian evolution, $\tilde G(k^{2})=1$ so  
the noise $\eta$ supplies \emph{every} four-momentum.  To prevent $\eta$ from creating 
$\phi$-particles out of the vacuum one must set $\tilde G(\mu^{2})=0$.   

\subsubsection{4.2a Feynman Diagrams}

	It is possible to use Feynman diagrams to visualize calculations because 
the density matrix (\ref{E45}) is expressed as the sum of unitary evolutions.  
Suppose at first that there are no Fermions.  The perturbation expansion of 
(\ref{E45}) provides two kinds of lines.  One is $\wr$, representing a $\phi$-particle 
of four-momentum $k^{\nu}$.  The other is --- representing the propagator 
$\tilde G(k^{2})$: when the integrals over $\eta$ are performed, only even powers of 
$\eta$ give a contribution, and its correlation function $G(x-x')$ acts like 
a propagator connecting two $\phi$-lines. 

	The unusual feature is that the Feynman diagrams represent 
a density matrix.  For example, the diagram showing the lowest order vacuum excitation of 
a single $\phi$-particle is  
$\wr\hspace{-.024in}_{\_}\hspace{-.02in}_{\_}\hspace{-.035in}\wr$.  
One must think of a vertical line down the middle of the diagram, separating it into the 
parts representing the terms to the left 
and right of the initial density matrix $\rho (T_{0})$.  
To the left (right), the unitary transformation is time-ordered (time-reverse-ordered), 
so one must think of the four-momentum coming in from the future to the past ($\downarrow$) 
to the right and going from the past to the future ($\uparrow$) to the left.  The only 
thing that can cross from right to left are G-propagator lines.  

	In the diagram $\wr\hspace{-.024in}_{\_}\hspace{-.02in}_{\_}\hspace{-.035in}\wr$, 
four-momentum is conserved because there is zero four-momentum in the initial 
vacuum state (below the lines) and zero four-momentum going out: $k^{\nu}$ 
goes in at the right, then it passes through the G-propagator and goes out at the left.  The 
density matrix associated with this diagram is 

\begin{eqnarray*}
\rho(T)=\gamma \tilde G(\mu^{2})T\int \frac{d{\bf k}}{\omega ({\bf k})}
a^{\dagger}({\bf k})|0><0|a({\bf k})
\end{eqnarray*}

\noindent ($a({\bf k})$ is the annihilation operator for the $\phi$-particle).  
The resulting energy increase is 
  
\begin{eqnarray*}
\mbox{Trace}\int d{\bf k}\omega ({\bf k})a^{\dagger}({\bf k})a({\bf k})\rho(T)
=\gamma \tilde G(\mu^{2})TV\frac{1}{(2\pi)^{3}}\int d{\bf k}\,.
\end{eqnarray*}
 
\noindent Thus, as mentioned earlier, if $\tilde G(\mu^{2})=0$ the vacuum excitation vanishes. However, 
this is only if there are no Fermions, and of course we must have Fermions since 
the whole point of collapse models is to collapse their states.

\subsubsection{4.2b Timelike Four-Momenta Cause Vacuum Excitation}\label{Timelike}
	
	When there are Fermions, then Fermion pairs can be excited from the vacuum since 
Fermions are coupled to the $\phi$-particles.   Pair production 
is represented by the diagram $\wr\hspace{-.024in}_{\_}\hspace{-.02in}_{\_}\hspace{-.035in}\wr$ 
with $\vee$ $\vee$ attached to the ends of the two $\phi$-lines. Since all four-momentum passes 
through the G-propagator, the diagram amplitude is proportional to $\tilde G[(p_{1}+p_{2})^{2}]$ 
where $p_{1}$, $p_{2}$ are the Fermion pair four-momenta.  
To make this vanish we must have $\tilde G(k^{2})$ vanish for 
$k^{2}\geq (2m)^{2}$.

	But, that's not the end of it. Pairs of $\phi$ particles are produced via 
the diagram $\wr\hspace{-.024in}_{\_}\hspace{-.02in}_{\_}\hspace{-.035in}\wr$ 
with $\wr\hspace{-.02in}_{\bigcirc}\hspace{-.06in}\wr$  $\wr\hspace{-.02in}_{\bigcirc}\hspace{-.06in}\wr$ 
on the ends of the $\phi$-lines ($_{\bigcirc}$ is a closed Fermion loop). To avoid this we must have 
$\tilde G(k^{2})=0$ for $k^{2}\geq (2\mu)^{2}=(2\mbox{eV})^{2}$.  That leaves very little of 
the timelike spectrum on which $\tilde G(k^{2})$ doesn't vanish.  Indeed, it must vanish there too.  
When the Fermion is charged, photon production from the vacuum (via the previous 
diagram where a single photon emerges from the closed Fermion loop) occurs 
unless $\tilde G(k^{2})=0$ for $k^{2}\geq 0$.

\subsubsection{4.2c Enter the Tachyon}\label{Tachyon}  

	By removing timelike four-momenta from the noise spectrum $\tilde G(k^{2})$, vacuum 
production of particles  
with timelike four-momenta  is avoided.  But, with only spacelike 
four-momenta remaining in the spectrum, the immediate question is 
whether the nonMarkovian evolution (\ref{E45}) describes collapse and gives CSL in 
the nonrelativistic limit.  The answer is  yes. \emph{Timelike four-momenta 
excite the vacuum and spacelike four-momenta  cause collapse.} 

	Consider the $c\rightarrow\infty$ limit of $G$:
	
\begin{eqnarray}
G({\bf x}-{\bf x}',t-t')&&=\frac{1}{(2\pi)^{4}}
\int d\omega d{\bf k}e^{i\omega(t-t')-i{\bf k}\cdot({\bf x}-{\bf x}')}
\tilde G(\frac{\omega ^{2}}{c^{2}}-{\bf k}^{2})\nonumber\\
&&\longrightarrow\delta (t-t')\frac{1}{(2\pi)^{3}}
\int d{\bf k}e^{-i{\bf k}\cdot({\bf x}-{\bf x}')}\tilde G(-{\bf k}^{2})\,.
\label{E46}
\end{eqnarray}

\noindent In this limit, $G$ is Markovian.  If we choose 
$\tilde G(k^{2})$ to be nonvanishing at only 
a single spectrum four-momentum $-\mu^{2}$:

\begin{eqnarray}
\tilde G(k^{2})=\delta (k_{0}^{2}-{\bf k}^{2}+\mu^{2})\label{E47}
\end{eqnarray}

\noindent then it follows from (\ref{E46}) that

\begin{eqnarray}
G(x-x')\longrightarrow\delta (t-t')\frac{1}{(2\pi)^{2}}
\frac{\sin [|{\bf x}-{\bf x}'|/a]}{|{\bf x}-{\bf x}'|}\,.\label{E48}
\end{eqnarray}

\noindent The nonrelativistic limit (\ref{E48}) is a fine 
replacement for the CSL Gaussian\cite{Weber}. (For the record, 
the exact expression for $G(x)$ is $-(8\pi^{2}ax)^{-1}N_{1}(x/a)$  for spacelike 
$x\equiv({\bf x}^{2}-x_{0}^{2})^{\frac{1}{2}}$ and 
$G(x)=-(4\pi^{3}ax)^{-1}K_{1}(x/a)$ for 
timelike $x\equiv(x_{0}^{2}-{\bf x}^{2})^{\frac{1}{2}}$, whose spacelike oscillations 
with period $a$ and timelike decay with time constant $a/c$ gives rise to 
(\ref{E48})).

	(\ref{E47}) is the spectrum of a tachyon of mass $\mu=1/a$.   
With this choice, the scalar field of mass $\mu$ is redundant: it 
isn't needed any longer to get the parameter $a$ into the model. 
One may remove $\phi$ from the evolution equation (\ref{E43}), replacing it with 
 $N(x)\equiv:\negthinspace\negthinspace\bar{\psi}(x)\psi(x)\negthinspace\negthinspace:$ 
 (then $\gamma$ is 
dimensionless) or the trace of the stress tensor 
$M(x)\equiv m:\negthinspace\negthinspace\bar{\psi}(x)\psi(x)\negthinspace\negthinspace:$ 
(then $\gamma$ has the 
dimensionsion of $G$, i.e., mass$^{-2}$). The vacuum pair production 
diagram $\vee\hspace{-.06in}\_\hspace{-.01in}\_
\hspace{-.01in}\_\hspace{-.05in}\vee$ vanishes because 
the spacelike four-momentum of the tachyon propagator (formerly called the G-propagator) 
cannot equal the timelike four-momentum of the fermion pair.

	If asked to propose a natural nonlocal relativistic structure, it is likely 
you would mention the tachyon. It is interesting that 
the need here for the tachyonic structure did not 
arise from any abstract desire for it but rather from the needs to 
eliminate vacuum excitation in lowest order and to obtain the 
nonrelativistic CSL limit.  Tachyons and collapse are two distinct theoretical 
structures requiring careful treatment lest superluminal communication 
and the difficulties of causal loops creep in: it is interesting that they are conjoined here.

\subsubsection{4.2d Collapse Rate and Energy Production Rate}\label{Results} 

	We now have a model with evolution equation
	
\begin{eqnarray}|\psi,T>_{w}={\cal T}e^{-\frac{1}{4\gamma}
\int_{T_{0}}^{T}dxdx'
	[w(x)-2\gamma N(x)] G(x-x')[w(x')-2\gamma N(x')]}
	|\psi,T_{0}>\label{E49}	
\end{eqnarray}	
	
\noindent(we have used $N$ rather than $M$ 
to facilitate comparison with CSL's Eq. (\ref{E32})) which is finite in 
lowest order, and we can calculate things. The 
associated density matrix in Fourier form is     

\begin{eqnarray}\rho (T)&&={\cal T}\int D\eta e^
{-2\gamma\int_{-\infty}^{\infty} dxdx'\eta(x)G^{-1}(x-x')\eta(x')}\nonumber\\
&&\qquad\cdot e^{-i2\gamma\int_{T_{0}}^{T}dx\eta(x)[N(x)\otimes 1-1\otimes N(x)]}\rho(T_{0})
\label{E50}\\
&&=\rho(T_{0})-\frac{\gamma}{2!}\int_{T_{0}}^{T}dxdx'G(x-x'){\cal T}[N(x),[N(x'),\rho(T_{0})]]+...
\label{E51}
\end{eqnarray}

\noindent CSL's comparable expressions are of the same form, where one replaces  
$G(x-x')$ by $\delta(t-t')\exp -(x-x')^{2}/4a^{2}$, and 
$N$ by the nonrelativistic particle number density operator.

	The Feynman diagrams corresponding to (\ref{E51}), 
describing collapse of a single Fermion in lowest order, 
are $\subset\hspace{-.035in}\mid$ $\mid$ and  
$\mid$ $\mid\hspace{-.035in}\supset$ which give the rate of depletion of  
the initial state (coming from (\ref{E51})'s 
$N(x)N(x')\rho(T_{0})$ and $\rho(T_{0})N(x)N(x')$ respectively) and
$|\hspace{-.06in}-\hspace{-.06in}|$ which gives the transition rate to 
the new state (coming from $N(x)\rho(T_{0})N(x')$). 
 
	In the following calculations I have replaced $N(x)=\bar{\psi}(x)\psi(x)$ 
with $N(x)=2m\psi^{2}(x)$ where $\psi$ is 
a Boson field to simplify the calculation by avoiding Dirac algebra.  
The relativistic energy increase of the particle is found to be 

\begin{eqnarray*}\frac{d}{dt}\bar H(t)=\frac{1}{2\pi^{2}}\gamma\frac{\mu^{3}}{m}
\sqrt{1+(\frac{\mu}{2m})^{2}}\end{eqnarray*}

\noindent which is independent of the particle's four-momentum ($d\bar H(t)/dt$ 
is an invariant). For $\mu/m<<1$ and 
$\gamma=\lambda \mu^{-1}$ this is the same (apart from numerical factors) as 
the nonrelativistic rate $3\lambda/8ma^{2}$ of section 3.1c.  

 	The relativistic collapse rate of a particle in an 
 initial state $\alpha|L>+\beta|R>$, where $|L>$ and $|R>$ represent two
 widely separated identical wavepackets with 
 momentum-space wavefunction $\Psi({\bf p})$, is characterized 
 by the decay of the off-diagonal density matrix element:
  
\begin{eqnarray*}\frac{d}{dt}<L|\rho(T)|R>=
-\alpha\beta^* \frac{1}{\pi^{2}}\gamma \mu
\sqrt{1+(\frac{\mu}{2m})^{2}}
\int d{\bf p}\frac{m}{E}|\Psi({\bf p})|^{2}\end{eqnarray*}

\noindent In the nonrelativistic limit one can set $E=m$ in the integrand and the result is the same 
as the CSL result $-\alpha\beta^*\lambda$ (see the equation following 
(\ref{E34}) in section 3.1a). More generally, if the particle has a fairly well 
defined momentum we may make the approximation 
$|\Psi({\bf p})|^{2}\approx \delta({\bf p}-{\bf p}_{0})$, with the result that 
the collapse rate, $\sim\sqrt{1-v_{0}^{2}}$, is time dilated.  
 
\subsection{ Finite RCSL Model for Free Particles}

 	Beyond lowest order there are diagrams with internal Fermion lines and, like the Hydra, 
 vacuum excitation rears its head again.  Pairs are produced from the vacuum 
 via the second order diagram $\vee\hspace{-.06in}\_\hspace{-.01in}\_\hspace{-.01in}\_
\hspace{-.14in}-\hspace{-.07in}\vee$.  

	Moreover, a free Fermion can gain infinite energy.  
From the first order diagram $|\hspace{-.07in}-\hspace{-.07in}|$ (utilized in the last section), 
where the incoming and outgoing lines are on-shell, easy 
kinematics shows that an incident Fermion at rest 
gains a small energy $\mu^{2}/2m$ ($\approx 10^{-6}$eV for an electron) and momentum 
$\mu\sqrt{1+(\mu/2m)^{2}}$ by emitting a (negative energy) tachyon.  
However, a Fermion can gain \emph{arbitrary} energy via the second order diagram 
$|\hspace{-.03in}^{-}_{-}\hspace{-.06in}^{-}_{-}\hspace{-.04in}|$. 

	There are two reasons for this.  
One is that the sum of four-momenta of two tachyons is capable of 
adding up to \emph{any} four momentum, e.g., the sum or difference of two Fermion four-momenta.  
The other is that a Feynman Fermion propagator $\sim (p^{2}- m^{2}+i\epsilon)^{-1}$ is 
capable of carrying away from a vertex \emph{any} four momentum, e.g., the sum or 
difference of a Fermion and tachyon four-momentum. 

	At present I know of three approaches to try solving this problem, 
only one of which 
I have spent some time at.  In one (largely unexplored) approach, the  
idea is to limit the four-momentum a particle propagator can transfer by 
introducing a relativistically invariant cutoff, effectively ``smearing" $N$ 
($:\negthinspace\negthinspace\bar{\psi}(x)\psi(x)\negthinspace\negthinspace:\rightarrow 
\int_{\sigma_{0}}^{\sigma}dx_{1}dx_{2}F(x,x_{1},x_{2})
:\negthinspace\negthinspace\bar{\psi}(x_{1})\psi(x_{2})\negthinspace\negthinspace:$).  
A second 
(unexplored) approach has been suggested by Diosi (private communication).  
He notes that the density matrix for electrodynamics may be traced over 
the electromagnetic field, resulting in a more general nonMarkovian density matrix  
than we have been using here for RCSL\cite{Diosi3}.  However, it is a form for which 
Strunz\cite{Strunz} has given a collapse dynamics (and for which a simpler collapse 
dynamics than Strunz's is given at the end of section 2.3e).  This statevector and density matrix 
evolution describes just the particle behavior in electrodynamics 
(as in Wheeler and Feynman's classical action-at-a-distance electrodynamics) and is 
relativistically invariant. The collapse dynamics 
replaces the decoherence due to radiation by particles. There is 
no energy creation problem---indeed, the extra $F$ term effectively absorbs energy, 
accounting for the radiative loss of electrodynamics. This is not a good candidate 
for a fundamental collapse model in that e.g., a massive body in its ground 
state in a superposition of \emph{here} plus \emph{there} will stay that way.  But, it 
suggests that it may be possible to construct a satisfactory collapse 
model with the extra freedom of a more general nonMarkovian density matrix. 

\subsubsection{4.3a Removing Time Ordering}

	The approach I \emph{shall} discuss is unconventional, but it may be simply stated.  
The evolution equation is (\ref{E49}) in Fourier form, except  \emph{with the  
time ordering removed}:  

\begin{eqnarray}
|\psi,T>_{w}&&=\int D\eta e^
{-2\gamma\int_{-\infty}^{\infty} dxdx'\eta (x) G^{-1}(x-x')\eta(x')}\nonumber\\
&&\qquad\qquad \cdot e^{i\int_{T_{0}}^{T}dx\eta(x)[w(x)-2\gamma N(x)]}|\psi,T_{0}>\,.
\label{E52}
\end{eqnarray}

\noindent Likewise, the associated density matrix is (\ref{E50}) with ${\cal T}$ removed.

	What does this achieve? The Feynman diagrams with or without time-ordering are the same.   
\emph{However}, the off-shell   
propagator $<0|{\cal T}\psi (x)\bar{\psi}(x')|0>$ is replaced by the on-shell  
propagator $<0|\psi (x)\bar{\psi}(x')|0>$.  That is, even when particle lines are 
internal, each vertex $|\hspace{-.03in}-\hspace{-.03in}$  describes a process 
where a real Fermion emits/absorbs a real tachyon, thereby gaining 
a limited amount of energy (it is readily shown that $\Delta E/E\leq\mu/m$ for $\mu<<m$).  
Moreover, there is no pair production, real or virtual, at any vertex: 
$\vee\hspace{-.06in}\_\hspace{-.01in}\_\hspace{-.01in}\_$ vanishes since 
a spacelike four-momentum cannot equal a timelike four-momentum:   

\begin{equation}\int_{-{T\over 2}}^{T\over 2} dx'G(x-x')\psi_{\pm}^{2} (x')\longrightarrow 0
\qquad \hbox {as\ }T\rightarrow \infty\label{E53}\end{equation}

\noindent ($\psi_{+}(x)$ is the positive frequency---the annihilation---part of $\psi(x)$).  The 
result is a \emph{completely finite S-matrix}: for large $T$, because of (\ref{E53}),

\begin{eqnarray}\rho (T/2)&&=\sum_{n=0}^{\infty} {(-4\gamma)^{n}\over (2n)!}
\int_{-T/2}^{T/2} dx_{1}...dx_{2n}
\sum_{C}G(x_{i_{1}}-x_{i_{2}})...G(x_{i_{2n-1}}-x_{i_{2n}})\nonumber\\
&&\qquad\qquad\cdot 
[\bar{\psi}_{-}(x_{1})\psi_{+}(x_{1}),[...[\bar{\psi}_{-}(x_{2n})
\psi_{+}(x_{2n}),\rho(-T/2)]...]]\label{E54}\end{eqnarray}

\noindent ($C$ means the sum is over all combinations of pairs of indices; 
for simplicity I have assumed that no antiparticles are initially present 
and so have omitted $\psi_{-}\bar{\psi}_{+}$ terms). 
We see from (\ref{E54}) that \emph{there is no need for renormalization}. One may also 
set $N(x)=\bar{\psi}_{-}(x)\psi_{+}(x)$ in the evolution equation (\ref{E52}).  

	Removing the time-ordering operation still gives an evolution which describes collapse.  
Indeed, in the usual test of collapse one sets $H=0$ and 
then, since operators are time-independent, time-ordering is irrelevant.

	It is worth remarking that removal of time ordering in a usual field 
theory results in the trivial S-matrix $S=1$ since energy-momentum 
conservation prevents a real process at each vertex such as a 
charged particle emitting a photon.  
It is also worth remarking that reinstating time-ordering in e.g., (\ref{E54}) would 
result in a nonrelativistic expression since ${\cal T}\psi_{+}(x)\bar{\psi}_{-}(x')$ is 
not Lorentz invariant  ($\psi_{+}(x)$, $\bar{\psi}_{-}(x')$ do not commute
at spacelike separation). A non-trivial finite relativistic S-matrix without time-ordering is 
possible because of the tachyon.

	This model, however, is only for free particles.  If the Fermion interacts 
with other particles, the free-particle field $\psi$ should be replaced 
by the Heisenberg field $\psi_{H}$.  The removal of time-ordering 
only applies to the $\psi_{H}$'s involved in the collapse interaction, not to 
those involved in e.g., the electromagnetic interaction, or else 
the usual electrodynamic physics will be destroyed.  But this allows 
the Feynman off-shell propagator to connect two vertices, one where a tachyon is 
emitted, the other where a photon is emitted, with the result that the 
photon can get infinite energy.

\subsubsection{4.3b Time-Ordering and No Time-Ordering}Removing  time-ordering 
introduces nonlocality.  How bad is it?

	To see what no time-ordering does, we begin by comparing the two 
perturbation series: 

\begin{eqnarray*}&&{\cal T}\{ e^{-i\int_{0}^{T}dtV(t)}\}=
\sum_{n}(-i)^{n}e^{iH_{0}T}\int_{0}^{T}dt_{n}\cdot\cdot\cdot
\int_{0}^{t_{3}}dt_{2}\int_{0}^{t_{2}}dt_{1}\\
&&\qquad\qquad
e^{-iH_{0}(T-t_{n})}V(0)\cdot\cdot\cdot 
e^{-iH_{0}(t_{3}-t_{2})}V(0)
e^{-iH_{0}(t_{2}-t_{1})}V(0)e^{-iH_{0}t_{1}}\\
\\
&&\ e^{-i\int_{0}^{T}dtV(t)}=
\sum_{n}\frac{(-i)^{n}}{n!}e^{iH_{0}T}\int_{0}^{T}dt_{n}
e^{-iH_{0}(T-t_{n})}V(0)e^{-iH_{0}t_{n}}\cdot\cdot\cdot\\
&&\cdot e^{iH_{0}T}\int_{0}^{T}dt_{2}
e^{-iH_{0}(T-t_{2})}V(0)e^{-iH_{0}t_{2}}e^{iH_{0}T}\int_{0}^{T}dt_{1}
e^{-iH_{0}(T-t_{1})}V(0)e^{-iH_{0}t_{1}}\end{eqnarray*}
  	
\noindent With time-ordering, particles evolve freely 
for time $t_{1}$, exchange momenta, evolve freely for time $t_{2}-t_{1}$, 
exchange momenta, etc., until time $T$. Without time-ordering, particles evolve freely 
for time $t_{1}$, exchange momenta, evolve freely for time $T-t_{1}$ and then 
\emph{evolve backwards in time} for time $T$, and then do it all over again.

	The hallmark of no time-ordering is backwards in time evolution.  Repeated 
forwards and backwards in time evolutions can take a particle 
outside its light cone.  However, as shall now be explained, in the present 
model this occurs with very small probabililty.

\subsubsection{4.3c Particle Evolution} Consider the wavefunction of a particle which starts out 
centered at ${\bf x}=0$ and with momentum centered at 
zero as well. Consider how this wavefunction spreads in lowest order. This turns out 
to be the same as the spread of a classical ensemble of particles.  
In this ensemble, each particle starts out at ${\bf x}=0$ with velocity ${\bf v}=0$.  
At random times in the interval $(0,T)$, particles 
suddenly emit a negative energy tachyon in a random direction, thereby 
gaining velocity $\approx(\mu/m)c$ (assuming $\mu<<m$).  Thereafter each particle 
continues moving freely forward in time for the rest of the time interval 
and, following this, it moves freely backward in time for $T$ sec.  

	A particle which received its 
impulse at $t=0$ travels a distance $(\mu/m)cT$ by time $T$ but, in its subsequent 
time-reversed evolution, returns to ${\bf x}=0$.  At the other extreme, a particle 
which receives its impulse at time $T$ ends up, after its time-reversed motion, 
at a distance $(\mu/m)cT$ from the origin in the direction opposite to its velocity.  
Thus the particles end up distributed in a sphere of radius $(\mu/m)cT$, with velocities 
of magnitude $(\mu/m)c$ directed radially inward. These distributions describe the wavefunction.   
(If the wavefunction is that of a particle initially moving with some particular nonzero velocity, 
the description is the appropriately Lorentz invariant one, e.g., the spread of 
the wavefunction is Lorentz contracted in the direction of motion.)

	If the particle emits/absorbs n tachyons, it undergoes 
n successive evolutions as described above.  The sum of ladder 
diagrams $|\hspace{-.05in}-\hspace{-.06in}|$ + 
$|\hspace{-.03in}^{-}_{-}\hspace{-.06in}^{-}_{-}\hspace{-.04in}|$ +... essentially 
describes the particle as undergoing a random walk with spread $\approx(\mu/m)cT$ (this spreads 
faster than the usual random walk $\sim\sqrt{T}$). The superluminal part of 
the wavefunction is negligibly small--- a tail.

\subsubsection{4.3d Approximate Locality}Consider two separate clumps of particles, 
L and R, at $t=0$, and the light cones emanating from them.  Assume the light cones do not 
overlap at time $t=T$.  Suppose the statevector evolution is unitary, 
$|\psi,T>=U(T)|\psi,0>$.  One may demonstrate locality, that anything going on 
in R during the interval $(0,T)$ doesn't affect what goes on in L,  
by showing that 
Trace$_{R}\{U(T)\rho(0)U^{\dagger}(T)\}=U_{L}'(T)$Trace$_{R}\{\rho(0)\}U_{L}'^{\dagger}(T)$,
where $U_{L}'$'s operators only act upon the particles in L. 

	In a standard quantum field theory, where the operators in $U$ are time-ordered, 
one may write $U(T)|\psi,0>=U_{R}'(T)U_{L}'(T)|\psi,0>$, where the operators in 
$U_{R}'(T)$ ($U_{R}'(T)$) act only within R's (L's) lightcone, and the above result follows. 
One can do this because the operators within R's lightcone commute with the operators 
within L's lightcone \emph{and} because the operators in $U$ which are not in these lightcones 
and which may not commute with R's or L's operators do not contribute to the evolution of 
$|\psi,T>$ (they act on the no-particle state in their region of spacetime and their 
contribution gets negligibly small for $T>>\hbar/2mc^{2}$, the virtual pair production time).  

	In the model presented here, the density matrix in Fourier form  
is a Gaussian-weighted superposition of unitary transformations $U_{\eta}$, 
where the operators in $U$ are not time-ordered. Because of the attendant forward 
and backward in time evolution, the operators in $U$ which act on e.g., R's 
particles, occupy more than R's lightcone.  \emph{Neglecting the tails}, R's particles 
at time $T$ are within R's lightcone, but the operators that 
got them there occupy a cylinder whose cross-section at any time is 
the cross-section of the lightcone at time $T$. While all the operators within 
R's cylinder do not commute with all the operators within L's cylinder, nonetheless 
the operators within R's cylinder do not act on L-particles, and vice versa.  
Therefore it is still true in this case that $U(T)|\psi,0>=U_{L}'(T)U_{R}'(T)|\psi,0>$, 
where the operators in $U_{R}'(T)$ ($U_{L}'(T)$) act only within R's (L's) cylinder, 
and the above-mentioned demonstration of locality (or rather approximate locality, since we 
have neglected the tails) goes through. 

	This shows that it is possible to make 
a finite relativistic collapse model.  As more such models are constructed, 
common features may be separable from model-specific ones.  This can lead to a better understanding 
of what is necessary in such models and, perhaps, a model with particularly 
satisfying features may stand out.       

\subsection{Concluding Remarks}

	There is a big difference between a conditional statement and an absolute statement: ``\emph{if} you 
win the lottery \emph{then} you will get ten million dollars" can't compare with ``you have 
won the lottery and you get ten million dollars."  

	The statements of SQT are conditional.  Faced with the statevector $c_{1}|a_{1}>+c_{2}|a_{2}>$, 
SQT says ``\emph{if} this is the description of a completed measurement \emph{then} the physical 
state is $|a_{1}>$ or $|a_{2}>$."  But actually, what the \emph{if} 
is conditioned upon, what the words ``a completed measurement" mean, lies outside the theory's ken.  SQT 
is not a complete description of nature because it fails to predict a physical phenomenon, 
namely that an event does---or does not---occur.  

	Phenomenological models are introduced into physics to describe phenomena 
that present theory fails to adequately treat.  Collapse models are phenomenological models.  
Their statements are absolute. Faced with the statevector $c_{1}|a_{1}>+c_{2}|a_{2}>$, the 
collapse model says that represents the physical state.  If you wait a bit, it may happen 
that the statevector is unchanged, and that's that.  Or, it may occur that the statevector rapidly 
evolves to $|a_{1}>$ or $|a_{2}>$, and again that's that. By ``that's that" is meant, in all cases, that  
the statevector represents the physical state: the model \emph{tells you} whether or not an 
event occurred.

	A phenomenological model becomes convincing when it produces experimentally 
verified predictions beyond those engineered into its 
construction and achieves respectability when it is 
wedded to the rest of physics. So far, neither of these has happened 
for collapse models, but they are fairly young.  However,  
the predictions are there (section 3.1c) and perhaps, in the hints 
of gravity (sections 3.1c,d) or tachyons (sections 4.2---in string theory models, 
tachyonic modes of strings appear---and are gotten rid of) there may be the seeds of 
a connection to another physics domain.             
    
\subsection*{Acknowledgements} I would like to thank the Istituto Italiano per gli Studi Filosofici 
and Heinz-Peter Breuer and Francesco Petruccione for initiating the 
conference and this book which grew out of it. I also wish to thank the 
Institute for Advanced Studies at the 
Hebrew University of Jerusalem and Yakir Aharonov for inviting me to participate in 
the workshop on Foundations of Physics where conversations with David Albert, Jeeva Anandan, 
Lajos Diosi, Andrew Frenkel, Dan Rohrlich, Abner Shimony, Jeff Tollakson, Bill Unruh, Lev Vaidman and 
Wojciech Zurek affected this work.

\end{document}